\documentclass[%
 reprint,
 longbibliography,
nofootinbib,
 amsmath,amssymb,
 aps, superscriptaddress,
pra,
]{revtex4-2}

\usepackage{graphicx}
\usepackage{dcolumn}
\usepackage{bm}
\usepackage{amsmath}
\usepackage{amssymb}
\usepackage{amsthm}
\usepackage{xcolor}

\usepackage{tikz}
\usepackage{tikz-3dplot}
\usetikzlibrary{arrows.meta}
\usepackage[caption=false]{subfig}
\usepackage{fontawesome5}
\usepackage{pgfplots}
\pgfplotsset{compat=1.18}

\newtheorem{theorem}{Theorem}
\newtheorem{proposition}{Proposition}
\newtheorem{corollary}{Corollary}
\newtheorem{lemma}{Lemma}
\newtheorem{definition}{Definition}
\newtheorem{remark}{Remark}
\newtheorem{example}{Example}
\newcommand{\Tr}{\operatorname{Tr}}
\newcommand{\bra}[1]{{\left\langle #1 \right|}}
\newcommand{\ket}[1]{{\left| #1 \right\rangle}}
\makeatletter
\newcommand{\braket}[1]{\@braket{#1}}
\def\@braket#1{\@ifnextchar\bgroup{\@@braket{#1}}{\@@braket{#1}{#1}}}
\def\@@braket#1#2{\left\langle #1 \middle| #2 \right\rangle}

\newcommand{\ketbra}[1]{\@ketbra{#1}}
\def\@ketbra#1{\@ifnextchar\bgroup{\@@ketbra{#1}}{\@@ketbra{#1}{#1}}}
\def\@@ketbra#1#2{\left| #1 \middle\rangle \middle\langle #2 \right|}
\makeatother

\begin{document}


\title{Quantum Advantage in Privacy-Preserving Hypothesis Testing via
Entanglement-Assisted Classical Communication}

\author{Ji-Ho Jung}
 \email{mkll321@kaist.ac.kr}
 \affiliation{
 School of Electrical Engineering, Korea Advanced Institute of Science and Technology (KAIST), 291 Daehak-ro, Yuseong-gu, 34141 Daejeon, Republic of Korea
}
\author{Seung-Hyun Nam}
 \email{seunghyun.sh.nam@jyu.fi}
\affiliation{
 Faculty of Information Technology, University of Jyväskylä, Mattilanniemi 2, 40100 Jyväskylä, Finland
}

\author{Si-Hyeon Lee}
\email{sihyeon@kaist.ac.kr}

\affiliation{
 School of Electrical Engineering, Korea Advanced Institute of Science and Technology (KAIST), 291 Daehak-ro, Yuseong-gu, 34141 Daejeon, Republic of Korea
}

\date{\today}
\begin{abstract}
Quantum mechanisms satisfying quantum local differential
privacy (QLDP) have been shown to achieve a quantum
advantage over classical local differential privacy (LDP)
mechanisms in certain hypothesis-testing problems.
While previous demonstrations have considered the
transmission of quantum states from users to a server,
we investigate whether this advantage can persist under
entanglement-assisted classical communication (EACC),
where users and the server share entanglement in advance
and communicate only classical messages.
We formulate an EACC-based hypothesis-testing framework
and introduce measurement local differential privacy
(MLDP) to protect private inputs against arbitrary
shared quantum states and server-side measurements.
We establish that QLDP classical-to-quantum channels admit MLDP-compliant EACC simulations and that, in the high-privacy regime, certain optimal QLDP mechanisms can be realized with one-bit classical communication, albeit using high-dimensional entanglement.  
We then demonstrate that a strict quantum advantage
persists even when the pre-shared entanglement is
restricted to a single qubit pair and each user
communicates only one classical bit.
Using EACC protocols based on the geometry of qubit
measurements, we establish this advantage for both
asymmetric and symmetric hypothesis testing and every
input alphabet of size $v \geq 3$ in the high-privacy limit.
The ratio of the achieved error exponent to the
optimal one under the LDP constraint is at least $5/4$ for all such alphabets and
approaches $3/2$ as $v$ grows.
These results demonstrate that quantum advantage in
locally private hypothesis testing can persist under
simultaneous constraints on entanglement dimension and
classical communication. 
\end{abstract}

\maketitle

\section{Introduction}
Statistical inference often relies on data collected from multiple
users, raising concerns about the disclosure of sensitive
individual information~\cite{evans2026,jegorova2023}.
Local differential privacy (LDP)~\cite{kasiviswanathan2011,duchi2013,kairouz2014}
provides a framework for protecting individual privacy without
requiring users to disclose their raw data to a server.
Under LDP, each user locally randomizes their input before transmission,
constraining how much the distribution of the reported output
can vary across different possible inputs.
This randomization limits the information available for inference, giving rise to the fundamental privacy–utility tradeoff (PUT).

A natural question is whether quantum resources can improve the
PUT beyond what is achievable with classical LDP.
One setting in which this question has been investigated is quantum
local differential privacy (QLDP)~\cite{yoshida2020,Yoshida_2025},
where each user encodes their private input into a quantum state
through a classical-to-quantum (CQ) channel and transmits the
resulting state to a server.
In this setting, QLDP constrains the distinguishability of the
quantum states associated with different inputs, providing a
quantum counterpart to the classical privacy guarantee.\footnote{
QLDP also refers to privacy-preserving quantum channels acting
on quantum inputs~\cite{zhou2017,hirche2023}, whereas we focus
on CQ preparation channels.}
Recent studies of privacy-preserving hypothesis testing have demonstrated that QLDP mechanisms can outperform classical LDP mechanisms for a certain class of hypotheses~\cite{nam2026,yoshida2026}.
In particular, the quantum advantage has been quantified
exactly in the high-privacy limit by characterizing and
comparing the optimal error exponents under classical
LDP and QLDP~\cite{yoshida2026}.

Previous demonstrations of quantum advantage in
privacy-preserving hypothesis testing under QLDP
have considered settings in which users transmit
quantum states to the server. This raises the question of whether direct quantum
communication is essential for such an advantage.
A complementary setting is entanglement-assisted classical communication (EACC), where users and
the server share entanglement in advance and
communicate only classical messages.
Quantum teleportation~\cite{gen_quan_telep}
and remote state
preparation~\cite{bennett2001rsp,hadiashar2018rsp,oufkir2024cqsim}
provide established methods for transferring or
remotely preparing quantum states using these
resources.\footnote{
Here, EACC refers to classical communication assisted by
pre-shared entanglement, rather than entanglement-assisted
classical communication over quantum
channels~\cite{bennett1999eacc}.}
Moreover, recent work has demonstrated quantum
advantage in certain communication tasks under
input-distinguishability constraints using only
a single entangled qubit pair and one bit of
classical communication~\cite{manna2026}.
These developments motivate us to investigate
whether quantum advantage in privacy-preserving
hypothesis testing can also be achieved under EACC,
and how stringent the constraints on entanglement
and classical communication can be while
preserving this advantage.
In particular, can such an advantage be achieved
with only a single entangled qubit pair and one
bit of classical communication per user?

To address these questions, we develop a framework for
privacy-preserving hypothesis testing based on EACC.
Each user and the server pre-share an entangled state
independent of the user's private input, after which the
user performs an input-dependent measurement on their
local subsystem and communicates the outcome to the server.
We introduce measurement local differential privacy (MLDP)
to impose a privacy constraint on these input-dependent
measurements.
Specifically, MLDP bounds the ratio of the joint
probabilities of the communicated outcome and any
subsequent server-side measurement outcome across
different private inputs, uniformly over all pre-shared
entangled states and server-side measurements.

We first show that any QLDP CQ 
channel can be simulated by an EACC protocol satisfying
MLDP, using generalized quantum
teleportation~\cite{gen_quan_telep}.
We then focus on the high-privacy regime and exploit
the structure of optimal QLDP mechanisms based on
Radon--Hurwitz equi-isoclinic tight fusion frames
(EITFFs)~\cite{yoshida2026} to reduce the classical
communication to a single bit per user while preserving
their optimal error exponents.
This result parallels the one-bit implementation of
optimal classical LDP mechanisms using pre-shared
classical randomness in the high-privacy
regime~\cite[Remark 4]{nam2024lowcom}.
However, although these EACC protocols require only
one bit of classical communication, the local dimension
of their pre-shared entangled states grows exponentially
with the input alphabet size.

We therefore investigate whether the entanglement
requirement can also be reduced to a single entangled
qubit pair without sacrificing the one-bit communication
constraint or the quantum advantage.
To this end, we construct MLDP-compliant EACC protocols
based on geometric arrangements of qubit measurements
on the Bloch sphere, supplemented by shared classical
randomness when needed.
We prove that these protocols achieve a strict quantum
advantage over optimal classical LDP mechanisms in the
high-privacy limit for every non-binary input alphabet.
Specifically, for the hypothesis-testing problems
considered here, the ratio of the achievable error
exponent to the optimal one under the LDP constraint 
is at least $5/4$ for every alphabet size $v \geq 3$
and approaches $3/2$ as $v$ grows.
Thus, quantum advantage persists with only a single
shared entangled qubit pair and one classical bit
per user.

The remainder of this paper is organized as follows.
Sec.~\ref{sec:preliminaries} reviews privacy-preserving
hypothesis testing and relevant classical and quantum results.
Sec.~\ref{sec:system_model} introduces the EACC framework
and measurement local differential privacy.
Sec.~\ref{sec:summary} summarizes our main results.
Sec.~\ref{sec:main} establishes the theoretical foundations
of EACC implementations and develops protocols that achieve
quantum advantage under restricted entanglement and
communication resources.
Sec.~\ref{sec:conclusion} concludes the paper.

\section{Hypothesis-Testing Framework and Preliminaries}\label{sec:preliminaries}

\subsection{Notation}

Throughout the paper, for a positive integer $m$, we write $[m]:=\{1,\ldots,m\}$.
For a quantum system $A$, $\mathcal H_A$ denotes its finite-dimensional Hilbert space, $\mathcal L(\mathcal H_A)$ denotes the space of linear operators on $\mathcal H_A$, and $\mathcal D(\mathcal H_A)$ denotes the set of density operators on $\mathcal H_A$.
A system label used as an operator subscript indicates the system on which it acts; for example, $\rho_A\in\mathcal D(\mathcal H_A)$ denotes a density operator on system $A$.
The identity operator on $\mathcal H_A$ is denoted by $I_A$; when we only need to specify the dimension $d=\dim\mathcal H_A$, we write $I_d$.

A positive operator-valued measure (POVM) on system $A$ with output alphabet $\mathcal Y$ is a collection $\{M_y\in\mathcal L(\mathcal H_A)\}_{y\in\mathcal Y}$ satisfying $M_y\ge0$ and $\sum_yM_y=I_A$.
A CQ channel $Q:\mathcal X\to\mathcal D(\mathcal H_B)$ is identified with its output family $Q=(Q_x)_{x\in\mathcal X}$, where $Q_x:=Q(x)$.
We use $(\cdot)^T$, $(\cdot)^*$, and $(\cdot)^\dagger$ to denote the transpose, complex conjugate, and conjugate transpose, respectively, where the first two are taken with respect to the computational basis.
We use $\|T\|_1:=\Tr\sqrt{T^\dagger T}$ for the trace norm, and all logarithms are natural.

For two $d$-dimensional systems $A$ and $S$ with fixed computational bases, define the maximally entangled state
\begin{equation}\label{eq:max_ent_state}
    \ket{\Phi_d^+}_{AS}:=\frac1{\sqrt d}\sum_{j=0}^{d-1}\ket{j}_A\ket{j}_S,\; \Phi_{AS}^d:=\ketbra{\Phi_d^+}_{AS}.
\end{equation}

\subsection{Hypothesis-Testing Scenarios} \label{subsec:hypothesis}
We consider a hypothesis-testing setting involving
$n$ users and a server.
Each user $i\in[n]$ holds a private datum $X_i$
taking values in a finite alphabet $\mathcal X=[v]$.
The data $X_1,\ldots,X_n$ are assumed to be
independently and identically distributed according
to an unknown underlying distribution $P$ on
$\mathcal X$.
The server aims to test hypotheses about the underlying distribution using information provided by the users.

Following~\cite{nam2026,yoshida2026}, we consider
the following family of smoothed point-mass
distributions on $\mathcal X$.
For a fixed $\eta\in(0,1]$, let
\begin{equation}\label{eq:smoothed_family}
    P^{\eta,h}(x)
    :=\eta\delta_{h,x}+\frac{1-\eta}{v},
    \qquad x,h\in[v],
\end{equation}
where $\delta_{h,x}$ denotes the Kronecker delta.
We also denote the uniform distribution by
$P^0(x):=1/v$.
Each $P^{\eta,h}$ assigns a higher probability
to the item $h$ than to any other item, with
the remaining items equally likely.

Two hypothesis-testing problems are defined as follows:
\begin{itemize}
    \item \emph{Asymmetric hypothesis testing:}
    Distinguish between the null hypothesis that
    the underlying distribution is $P^{\eta,h}$
    for some $h\in[v]$ and the alternative
    hypothesis that it is the uniform
    distribution $P^0$.

    \item \emph{Symmetric hypothesis testing:}
    Assuming that $P=P^{\eta,h}$ for some unknown
    $h\in[v]$, identify the most probable item $h$.
\end{itemize}

\subsection{Privacy-Preserving Hypothesis Testing Based on CQ Channels}\label{subsec:hypothesis_testing}
\begin{figure}
    \centering
    {\resizebox{\columnwidth}{!}{\begin{tikzpicture}[
    font=\rmfamily,
    >=Latex,
    iconnode/.style={align=center, inner sep=0pt},
    arrow/.style={->, thick, draw=cyan!70!black,
                  shorten >=2.5pt, shorten <=2pt},
    sourcearrow/.style={->, thick, draw=cyan!70!black,
                        shorten >=2.5pt, shorten <=0pt},
    desc/.style={align=center, font=\large, anchor=north},
    statelabel/.style={align=center, font=\large, anchor=north}
]

\def\yone{3.0}
\def\ytwo{0.0}
\def\yn{-3.9}
\def\yvdots{-2.0}
\def\ycenter{-0.45}
\def\yout{-0.70}
\def\xdist{-3.15}
\def\xbranch{-1.65}
\def\xuser{0}
\def\xenc{2.3}
\def\xatom{4.55}     
\def\xserver{7.20}
\def\xout{9.75}

\node[iconnode] (dist) at (\xdist,\ycenter) {\LARGE $P^h$};
\node[desc] at (\xdist,\ycenter-0.45)
    {Underlying\\distribution};
\coordinate (branch) at (\xbranch,\ycenter);

\node[draw, thick, rounded corners=3pt,
      minimum height=8.5cm, minimum width=2.25cm,
      fill=gray!5] (serverbox) at (\xserver,-0.35) {};
\node[iconnode, fill=white, inner sep=2pt]
    at (serverbox.north) {\Huge\faDatabase};
\node[align=center, font=\large] at (\xserver,\yout)
    {Server's\\measure-\\ment};

\node[iconnode] (decision) at (\xout,\yout)
    {\LARGE $\hat h$};
\node[desc] at (\xout,\yout-0.48)
    {Hypothesis\\decision};

\node[iconnode] (user1) at (\xuser,\yone) {\Huge\faUser};
\node[desc] at (\xuser,\yone-0.65) {User 1\\Input $x_1$};
\node[iconnode] (enc1) at (\xenc,\yone) {\Huge\faDesktop};
\node[fill=white, minimum width=18pt, minimum height=13pt,
      yshift=2pt] at (\xenc,\yone) {};
\node[yshift=2pt] at (\xenc,\yone) {\large\faAtom};
\node[desc] at (\xenc,\yone-0.68) {CQ encoder 1};

\node[iconnode] (user2) at (\xuser,\ytwo) {\Huge\faUser};
\node[desc] at (\xuser,\ytwo-0.65) {User 2\\Input $x_2$};
\node[iconnode] (enc2) at (\xenc,\ytwo) {\Huge\faDesktop};
\node[fill=white, minimum width=18pt, minimum height=13pt,
      yshift=2pt] at (\xenc,\ytwo) {};
\node[yshift=2pt] at (\xenc,\ytwo) {\large\faAtom};
\node[desc] at (\xenc,\ytwo-0.68) {CQ encoder 2};

\node[iconnode] at (\xuser,\yvdots) {\Large $\vdots$};
\node[iconnode] at (\xenc,\yvdots) {\Large $\vdots$};
\node[iconnode] at (\xatom,\yvdots) {\Large $\vdots$};

\node[iconnode] (usern) at (\xuser,\yn) {\Huge\faUser};
\node[desc] at (\xuser,\yn-0.65) {User $n$\\Input $x_n$};
\node[iconnode] (encn) at (\xenc,\yn) {\Huge\faDesktop};
\node[fill=white, minimum width=18pt, minimum height=13pt,
      yshift=2pt] at (\xenc,\yn) {};
\node[yshift=2pt] at (\xenc,\yn) {\large\faAtom};
\node[desc] at (\xenc,\yn-0.68) {CQ encoder $n$};

\draw[sourcearrow] (dist.east) -- (branch) -- (user2.west);
\draw[sourcearrow] (branch) -- (user1.west);
\draw[sourcearrow] (branch) -- (usern.west);

\draw[arrow] (user1.east) -- (enc1.west);
\draw[arrow] (user2.east) -- (enc2.west);
\draw[arrow] (usern.east) -- (encn.west);

\coordinate (srv1) at (serverbox.west |- enc1);
\coordinate (srv2) at (serverbox.west |- enc2);
\coordinate (srvn) at (serverbox.west |- encn);
\draw[arrow] (enc1.east) -- (srv1);
\draw[arrow] (enc2.east) -- (srv2);
\draw[arrow] (encn.east) -- (srvn);
\node[iconnode, fill=white, inner sep=2pt]
    at (\xatom,\yone) {\huge\faAtom};
\node[iconnode, fill=white, inner sep=2pt]
    at (\xatom,\ytwo) {\huge\faAtom};
\node[iconnode, fill=white, inner sep=2pt]
    at (\xatom,\yn) {\huge\faAtom};
\node[statelabel] at (\xatom,\yone-0.38)
    {Quantum\\state $Q_{x_1}$};
\node[statelabel] at (\xatom,\ytwo-0.38)
    {Quantum\\state $Q_{x_2}$};
\node[statelabel] at (\xatom,\yn-0.38)
    {Quantum\\state $Q_{x_n}$};

\coordinate (outputport) at (serverbox.east |- decision);
\draw[arrow] (outputport) -- (decision.west);
\end{tikzpicture}}}
    \caption{Privacy-preserving hypothesis testing
    based on CQ channels.}
    \label{fig:CQ}
\end{figure}
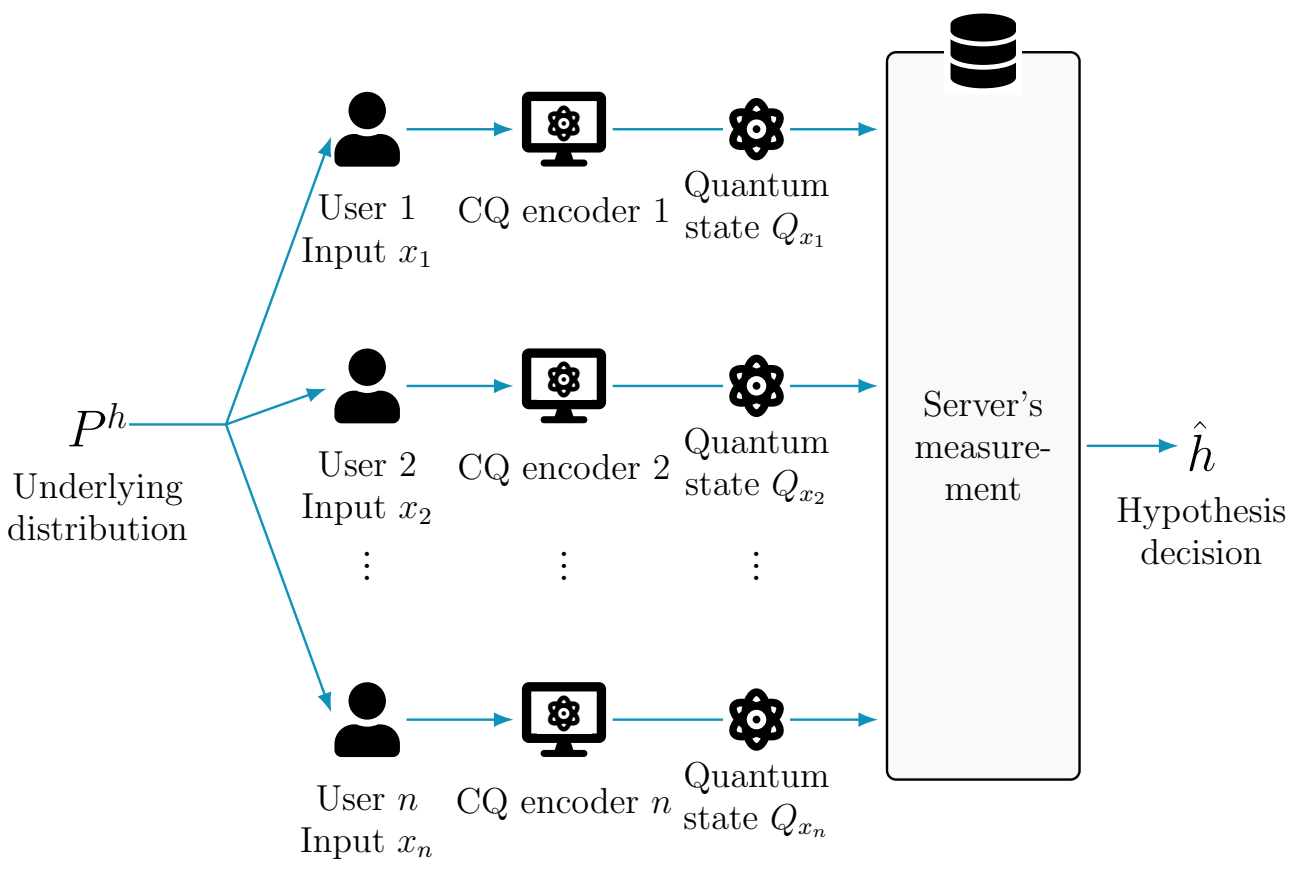

We  review privacy-preserving hypothesis testing
based on classical-to-quantum (CQ) channels, together
with its classical counterpart.
As illustrated in Fig.~\ref{fig:CQ}, each user
encodes their private datum $X_i$ into a quantum
state using a CQ channel
$Q:\mathcal X\to\mathcal D(\mathcal H_B)$
and transmits the state to the server.
To protect the private input, the channel is
required to satisfy the $\epsilon$-quantum local
differential privacy ($\epsilon$-QLDP) constraint
for a fixed $\epsilon>0$.

\begin{definition}[Quantum local differential
privacy~\cite{yoshida2020,Yoshida_2025}]
\label{def:qldp}
For $\epsilon>0$, a CQ channel
$Q:\mathcal X\to\mathcal D(\mathcal H_B)$
satisfies $\epsilon$-quantum local differential
privacy ($\epsilon$-QLDP) if
\begin{equation}\label{eq:qldp}
    Q_x\le e^\epsilon Q_{x'},
    \qquad
    \forall x,x'\in\mathcal X.
\end{equation}
In short, we refer to such a CQ channel $Q$
as an $\epsilon$-QLDP channel.
\end{definition}

Operationally, the $\epsilon$-QLDP constraint
guarantees that no measurement performed on the
output quantum state can yield a classical outcome
whose likelihood varies by more than a factor of
$e^\epsilon$ across any two distinct classical inputs.
Hence, $\epsilon$ can be interpreted as an upper
bound on the leakage of private information.

The server performs hypothesis testing by measuring
the quantum states received from the users.
Under the distributions $P^{\eta,h}$ and $P^0$
defined in Sec.~\ref{subsec:hypothesis}, the corresponding
single-user averaged output states are
\begin{align}\label{eq:omega_cq}
    \omega_{\eta,h}^{\mathrm{CQ}}(Q)
    &:=\sum_{x\in[v]}P^{\eta,h}(x)Q_x,\\
    \omega_0^{\mathrm{CQ}}(Q)
    &:=\sum_{x\in[v]}P^0(x)Q_x.
\end{align}
Since the users' data are independently and
identically distributed, the server receives
$[\omega_{\eta,h}^{\mathrm{CQ}}(Q)]^{\otimes n}$
under $P^{\eta,h}$, or
$[\omega_0^{\mathrm{CQ}}(Q)]^{\otimes n}$
under $P^0$.
When the channel is clear from the context,
we abbreviate these states as $\omega_{\eta,h}$
and $\omega_0$, respectively.

We now formulate the asymmetric and symmetric
hypothesis-testing problems introduced in
Sec.~\ref{subsec:hypothesis} in terms of these output states
and their asymptotic error exponents.

\subsubsection{Asymmetric Hypothesis Testing}

For asymmetric hypothesis testing, the server
performs a binary POVM $\{M,I-M\}$ on the
received quantum state to distinguish between
the null and alternative hypotheses defined
in Sec.~\ref{subsec:hypothesis}.
The outcome associated with $M$ corresponds
to accepting the null hypothesis, while the
outcome associated with $I-M$ corresponds
to accepting the alternative hypothesis.

The type-I and type-II error probabilities are defined as
\begin{align}
    \alpha^{(n)}_\eta\left(M\right)
    &:=\max_{h\in [v]}
    \Tr\left[(I-M)\omega_{\eta, h}^{\otimes n}\right],\\
    \beta^{(n)} (M)
    &:=\Tr\left[M\omega_0^{\otimes n}\right],
\end{align}
respectively.
In asymmetric hypothesis testing, we often consider the asymptotic type-II error exponent
\begin{multline}
    A\left(\{\omega_{\eta, h}\}_{h\in [v]},\omega_0\right) 
    \\ :=\liminf_{n\to\infty}
    -\frac1n \log \beta_\delta^{(n)}\left(\{\omega_{\eta, h}\}_{h\in [v]},\omega_0\right), 
\end{multline}
where, for $\delta \in (0,1)$,
\begin{multline}
    \beta_\delta^{(n)}\left(\{\omega_{\eta, h}\}_{h\in [v]},\omega_0\right)
    \\ :=\inf_{0\le M\le I}\left\{ 
        \beta^{(n)}(M):
    \alpha^{(n)}_\eta\left(M\right)\le\delta
    \right\}.
\end{multline}
By the composite quantum Stein's lemma~\cite{ogawa2000, Bjelakovi2005, hiai_proper_1991}, we obtain 
\begin{equation}\label{eq:asym_stein}
    A\left(\{\omega_{\eta, h}\}_{h\in [v]},\omega_0\right) = \min\limits_{h \in [v]} D(\omega_{\eta, h} \Vert \omega_0),
\end{equation}
where $D(\cdot\Vert\cdot)$ denotes the quantum relative entropy 
\begin{equation}
    \!\!\! D(\rho\Vert\sigma) := 
    \begin{cases}
        \frac{\Tr[\rho(\log\rho-\log\sigma)]}{\Tr \rho},\!\!\!
        &\operatorname{supp}\rho\subseteq\operatorname{supp}\sigma,\\
        +\infty,
        &\text{otherwise},
    \end{cases}
\end{equation}
for positive semidefinite $\rho$ and $\sigma$ with $\rho \ne 0$.

\subsubsection{Symmetric Hypothesis Testing}

For symmetric hypothesis testing, we assume
that the unknown index $h\in[v]$ follows a
prior distribution $\gamma$ with $\gamma(h)>0$
for all $h\in[v]$.
The server performs a POVM $\{M_k\}_{k\in[v]}$
on the received quantum state
$\omega_{\eta,h}^{\otimes n}$, where the outcome
associated with $M_k$ corresponds to identifying
$k$ as the most probable item.
Then, the minimum average error probability is given as
\begin{multline}
    P_e^{(n)}\left(\{\omega_{\eta, h}\}_{h\in[v]},\gamma\right)
    \\:=    \inf_{\{M_h\}}
    \sum_{h\in[v]}
    \gamma(h)
    \Tr\left[(I-M_h)\omega_{\eta, h}^{\otimes n}\right].
\end{multline}
Its asymptotic error exponent is defined by
\begin{align}
   \!\!\!\!\! S\left(\{\omega_{\eta, h}\}_{h\in[v]}\right)
    &:=\liminf_{n\to\infty}
    -\frac1n
    \log P_e^{(n)}\left(\{\omega_{\eta, h}\}_{h\in[v]},\gamma\right)
    \nonumber\\
    &=\min_{h\ne h'}\left[
    -\log
    \min_{s\in[0,1]}
    \Tr\left(\omega_{\eta, h}^s\omega_{\eta,h'}^{1-s}\right)
    \right], \label{eq:sym_Chernoff}
\end{align}
where the second equality follows from the multiple quantum Chernoff bound~\cite{nussbaum2011multiple, Li_2016_defquantumchernoff}.
This exponent is independent of the prior $\gamma$ as long as every hypothesis has positive prior probability.

\begin{remark}
    In the classical counterpart, each user utilizes a classical channel (conditional probability) $q(y|x)$ to encode their data. 
    Note that a conditional distribution can be considered as a CQ channel
    \begin{equation}\label{eq:diagonal_cq}
        Q_x^q:=\sum_{y\in\mathcal Y}q(y|x)\ketbra{y},
    \end{equation}
    and this identification, combined with the definition of $\epsilon$-QLDP, exactly recovers the definition of $\epsilon$-local differential privacy (LDP) for conditional probabilities~\cite{dwork2006,duchi2013}.
    
    In the classical scenario, under $P^{\eta,h}$ for $h\in[v]$, the server receives the classical state $[\omega_{\eta, h}^{\mathrm C}(q)]^{\otimes n}$, where 
    \begin{equation}\label{eq:omega_c}
        \omega_{\eta, h}^{\mathrm C}(q)
        :=\sum_{y\in\mathcal Y}
        \left(\sum_{x\in\mathcal X}P^{\eta, h}(x)q(y|x)\right)
        \ketbra{y}.
    \end{equation}
    The corresponding state under $P^0$, denoted by $\omega_0^{\mathrm C}(q)$, is defined analogously.
    The server then applies classical post-processing to perform the hypothesis test.
    For the two hypothesis testing tasks considered in this work, the asymptotic error exponents in the classical scenario~\cite{leang1997, levitan2002} are also given by Eqs.~\eqref{eq:asym_stein} and~\eqref{eq:sym_Chernoff} under the identification given in~\eqref{eq:diagonal_cq}.
    
    To encompass both the classical and quantum scenarios with a common notation, and to emphasize the dependence of the error exponents on the channel, we adopt the following notation throughout the remainder of this paper:
    \begin{align}\label{eq:asym_mech}
        A_\eta\left(Q\right)
        &:=
        A\left(
        \{\omega_{\eta, h}(Q)\}_{h\in [v]},
        \omega_0(Q)
        \right),\\
        S_\eta\left(Q\right)
        &:=
        S\left(
        \{\omega_{\eta, h}(Q)\}_{h\in[v]}
        \right),
        \label{eq:sym_mech}
    \end{align}
    where $Q$ can be either a CQ or a classical channel, with the classical case understood as in Eq.~\eqref{eq:diagonal_cq}.
\end{remark}

\subsubsection{Privacy--Utility Tradeoffs and Quantum Advantage}

We now review the optimal error exponents
achievable under the QLDP constraint and
compare them with their classical LDP counterparts.
For each hypothesis-testing task, the optimal
error exponent characterizes the corresponding
privacy--utility tradeoff (PUT).
Specifically, let $\mathfrak M_\epsilon^{\mathrm{CQ}}(d)$ be the set of all $\epsilon$-QLDP channels from $\mathcal{X}$ to $\mathcal D(\mathcal H)$, where $\mathcal H$ is a $d$-dimensional Hilbert space, and let $\mathfrak M_\epsilon^{\mathrm{CQ}}:=\bigcup_{d \in \mathbb{N}} \mathfrak M_\epsilon^{\mathrm{CQ}}(d)$. 
Then, the PUTs for asymmetric and symmetric hypothesis testing under the $\epsilon$-QLDP constraint are defined as
\begin{align}
    A_\eta[\mathfrak M_\epsilon^{\mathrm{CQ}}]
    &:=
    \sup_{Q\in\mathfrak M_\epsilon^{\mathrm{CQ}}}
    A_\eta\left(Q\right),
    \\S_\eta[\mathfrak M_\epsilon^{\mathrm{CQ}}]
    &:=
    \sup_{Q\in\mathfrak M_\epsilon^{\mathrm{CQ}}}
    S_\eta\left(Q\right),
\end{align}
respectively.
For the classical counterpart, we analogously define the set of all $\epsilon$-LDP channels $\mathfrak M_\epsilon^{\mathrm C}$ and the corresponding PUTs $A_\eta[\mathfrak M_\epsilon^{\mathrm C}]$ and $S_\eta[\mathfrak M_\epsilon^{\mathrm C}]$.
We call a channel optimal for a given hypothesis-testing task if it achieves the corresponding PUT.
We say that there exists a quantum advantage if $A_\eta[\mathfrak M_\epsilon^{\mathrm{CQ}}] > A_\eta[\mathfrak M_\epsilon^{\mathrm{C}}]$ or $S_\eta[\mathfrak M_\epsilon^{\mathrm{CQ}}] > S_\eta[\mathfrak M_\epsilon^{\mathrm{C}}]$.
For brevity, we refer to the PUTs under the $\epsilon$-QLDP and $\epsilon$-LDP constraints as the CQ PUTs and classical PUTs, respectively.

Regarding the quantum advantage, the results in~\cite{yoshida2020} imply that no quantum advantage exists when $v = 2$.
Subsequently, the work~\cite{nam2026} demonstrated a quantum advantage for certain ranges of $v$, $\eta$, and $\epsilon \lesssim 1$. 
A more recent study~\cite{yoshida2026} extended this advantage to all $v \geq 3$ and $\eta \in (0,1]$, while also characterizing the CQ PUTs and the corresponding optimal QLDP channels in the high-privacy regime where $\epsilon \approx 0$.

\begin{theorem}[{\cite[Thm.~3.3]{yoshida2026}}]
\label{thm:known_optima}
For fixed $v\ge2$ and $\eta\in(0,1]$,
the optimal PUTs for asymmetric and symmetric
hypothesis testing satisfy
\begin{align}
    A_\eta[\mathfrak M_\epsilon^{\mathrm{CQ}}]
    &=
    \frac{v-1}{4v}\eta^2\epsilon^2
    +o(\eta^2\epsilon^2),\\
    A_\eta[\mathfrak M_\epsilon^{\mathrm C}]
    &=
    \frac{\lfloor v/2\rfloor\lceil v/2\rceil}
    {2v^2}\eta^2\epsilon^2
    +o(\eta^2\epsilon^2),\\
    S_\eta[\mathfrak M_\epsilon^{\mathrm{CQ}}]
    &=
    \frac18\eta^2\epsilon^2
    +o(\eta^2\epsilon^2),\\
    S_\eta[\mathfrak M_\epsilon^{\mathrm C}]
    &=
    \frac{\lfloor v/2\rfloor\lceil v/2\rceil}
    {4v(v-1)}\eta^2\epsilon^2
    +o(\eta^2\epsilon^2),
\end{align}
as $\epsilon\to0$.
Consequently, for $v\ge3$,
\begin{multline}\label{eq:optimal_qc_ratio}
    \lim_{\epsilon\to0}
    \frac{A_\eta[\mathfrak M_\epsilon^{\mathrm{CQ}}]}
    {A_\eta[\mathfrak M_\epsilon^{\mathrm C}]}
    =
    \lim_{\epsilon\to0}
    \frac{S_\eta[\mathfrak M_\epsilon^{\mathrm{CQ}}]}
    {S_\eta[\mathfrak M_\epsilon^{\mathrm C}]}
    =
    \frac{v(v-1)}
    {2\lfloor v/2\rfloor\lceil v/2\rceil}
    \ge\frac32.
\end{multline}
Moreover, the optimal CQ PUTs can be
asymptotically achieved as $\epsilon\to0$
with a finite output dimension $d$ satisfying
\begin{equation}\label{eq:cq_sufficient_dimension}
    d\ge
    2^{\max\{1,\lceil(v-2)/2\rceil\}}.
\end{equation}
Specifically, for any such $d$,
\begin{equation}\label{eq:cq_dimension_asymptotic}
    \lim_{\epsilon\to0}
    \frac{A_\eta[\mathfrak M_\epsilon^{\mathrm{CQ}}(d)]}
    {A_\eta[\mathfrak M_\epsilon^{\mathrm{CQ}}]}
    =
    \lim_{\epsilon\to0}
    \frac{S_\eta[\mathfrak M_\epsilon^{\mathrm{CQ}}(d)]}
    {S_\eta[\mathfrak M_\epsilon^{\mathrm{CQ}}]}
    =1.
\end{equation}
\end{theorem}

We next introduce the isoclinic mechanisms
constructed from Radon--Hurwitz EITFFs,
which asymptotically achieve the optimal
CQ PUTs in Theorem~\ref{thm:known_optima}.

\begin{definition}[Equi-isoclinic tight fusion frame]
\label{def:eitff}
A collection of $v$ rank-$r$ orthogonal
projections $\{F_x\}_{x\in[v]}$ on a
$d$-dimensional Hilbert space is an
$\mathrm{EITFF}(d,r,v)$ if
\begin{align}
    \sum_{x=1}^v F_x
    &=
    \frac{vr}{d}I_d,\\
    F_{x'}F_xF_{x'}
    &=
    \frac{vr-d}{d(v-1)}F_{x'},
    \qquad x\ne x'.
\end{align}
\end{definition}

\begin{definition}[Isoclinic mechanism]
\label{def:isoclinic_mechanism}
Given an $\mathrm{EITFF}(d,r,v)$,
the associated $\epsilon$-QLDP isoclinic
mechanism is defined by
\begin{equation}\label{eq:isoclinic_mechanism}
    Q_x^{\mathrm{iso}}
    :=
    \frac{\mu_{\mathrm{iso}}}{d}I_d
    +
    \frac{1-\mu_{\mathrm{iso}}}{r}F_x,
    \qquad x\in[v],
\end{equation}
where $\mu_{\mathrm{iso}}$ is the smallest
real value for which $Q^{\mathrm{iso}}$
satisfies the $\epsilon$-QLDP constraint.
\end{definition}

Radon--Hurwitz EITFFs form a family of
$\mathrm{EITFF}(2r,r,v)$ that exists
if and only if
\begin{equation}
    v\le N_{\mathrm{RH}}(r)+2,
\end{equation}
where $N_{\mathrm{RH}}(r):=2a+2$ for
$r=2^a b$, with $a$ a nonnegative integer
and $b$ a positive odd integer.
The minimum dimension of such an EITFF is
\begin{equation}
    d=2^{\max\{1,\lceil(v-2)/2\rceil\}},
\end{equation}
which grows exponentially with $v$.

Having reviewed the quantum advantage
achievable with CQ channels, we now
investigate whether this advantage can
also be attained using entanglement-assisted
classical communication (EACC), where
users and the server share entanglement
in advance and communicate only classical
messages.

\section{Privacy-Preserving Hypothesis Testing Based on EACC}\label{sec:system_model}

\begin{figure}[t!]
    \centering
    {\resizebox{\columnwidth}{!}{
\begin{tikzpicture}[
    font=\rmfamily,
    >=Latex,
    iconnode/.style={align=center, inner sep=0pt},
    arrow/.style={->, thick, draw=cyan!70!black,
                  shorten >=2.5pt, shorten <=2pt},
    sourcearrow/.style={->, thick, draw=cyan!70!black,
                        shorten >=2.5pt, shorten <=0pt},
    entlink/.style={thick, densely dashed, draw=black!70},
    entatom/.style={iconnode, fill=white, inner sep=1.3pt},
    desc/.style={align=center, font=\large, anchor=north},
    msgdesc/.style={align=center, font=\large, anchor=north}
]

\def\yone{3.0}
\def\ytwo{0.0}
\def\yn{-3.9}
\def\yvdots{-2.0}
\def\ycenter{-0.45}
\def\yout{-0.70} 
\def\xdist{-3.15}
\def\xbranch{-1.65}
\def\xuser{0}
\def\xenc{2.2}
\def\xatom{4.55} 
\def\xserver{7.20}
\def\xout{9.75}

\node[iconnode] (dist) at (\xdist,\ycenter) {\LARGE $P^h$};
\node[desc] at (\xdist,\ycenter-0.45) {Underlying\\distribution};
\coordinate (branch) at (\xbranch,\ycenter);

\node[draw, thick, rounded corners=3pt,
      minimum height=8.5cm, minimum width=2.25cm,
      fill=gray!5] (serverbox) at (\xserver,-0.35) {};
\node[iconnode, fill=white, inner sep=2pt]
      at (serverbox.north) {\Huge\faDatabase};
\node[align=center, font=\large] at (\xserver,\yout)
      {Server's\\measure-\\ment};

\node[iconnode] (decision) at (\xout,\yout) {\LARGE $\hat h$};
\node[desc] at (\xout,\yout-0.48) {Hypothesis\\decision};

\node[iconnode] (user1) at (\xuser,\yone) {\Huge\faUser};
\node[desc] at (\xuser,\yone-0.65) {User 1\\Input $x_1$};
\node[iconnode] (enc1) at (\xenc,\yone) {\Huge\faDesktop};
\node[fill=white, minimum width=18pt, minimum height=13pt,
      yshift=2pt] at (\xenc,\yone) {};
\node[yshift=2pt] at (\xenc,\yone) {\large\faIcon{tachometer-alt}};
\node[desc] at (\xenc,\yone-0.68)
      {$x_1$-dependent\\POVM $\Lambda_{y_1|x_1}$};

\node[iconnode] (user2) at (\xuser,\ytwo) {\Huge\faUser};
\node[desc] at (\xuser,\ytwo-0.65) {User 2\\Input $x_2$};
\node[iconnode] (enc2) at (\xenc,\ytwo) {\Huge\faDesktop};
\node[fill=white, minimum width=18pt, minimum height=13pt,
      yshift=2pt] at (\xenc,\ytwo) {};
\node[yshift=2pt] at (\xenc,\ytwo) {\large\faIcon{tachometer-alt}};
\node[desc] at (\xenc,\ytwo-0.68)
      {$x_2$-dependent\\POVM $\Lambda_{y_2|x_2}$};

\node[iconnode] at (\xuser,\yvdots) {\Large $\vdots$};
\node[iconnode] at (\xenc,\yvdots) {\Large $\vdots$};
\node[iconnode] at (\xatom,\yvdots) {\Large $\vdots$};

\node[iconnode] (usern) at (\xuser,\yn) {\Huge\faUser};
\node[desc] at (\xuser,\yn-0.65) {User $n$\\Input $x_n$};
\node[iconnode] (encn) at (\xenc,\yn) {\Huge\faDesktop};
\node[fill=white, minimum width=18pt, minimum height=13pt,
      yshift=2pt] at (\xenc,\yn) {};
\node[yshift=2pt] at (\xenc,\yn) {\large\faIcon{tachometer-alt}};
\node[desc] at (\xenc,\yn-0.68)
      {$x_n$-dependent\\POVM $\Lambda_{y_n|x_n}$};

\draw[sourcearrow] (dist.east) -- (branch) -- (user2.west);
\draw[sourcearrow] (branch) -- (user1.west);
\draw[sourcearrow] (branch) -- (usern.west);

\draw[arrow] (user1.east) -- (enc1.west);
\draw[arrow] (user2.east) -- (enc2.west);
\draw[arrow] (usern.east) -- (encn.west);

\coordinate (srv1) at (serverbox.west |- enc1);
\coordinate (srv2) at (serverbox.west |- enc2);
\coordinate (srvn) at (serverbox.west |- encn);
\coordinate (entaxis) at (\xatom,0);

\draw[arrow] (enc1.east) -- (srv1)
    coordinate[midway] (msgbase1);

\draw[arrow] (enc2.east) -- (srv2)
    coordinate[midway] (msgbase2);

\draw[arrow] (encn.east) -- (srvn)
    coordinate[midway] (msgbasen);

\coordinate (msgpos1) at
    ([xshift=0pt, yshift=0pt]msgbase1);

\coordinate (msgpos2) at
    ([xshift=0pt, yshift=0pt]msgbase2);

\coordinate (msgposn) at
    ([xshift=0pt, yshift=0pt]msgbasen);


\node[msgdesc, xshift=6pt, yshift=-1pt]
    (msg1) at (msgpos1)
    {{\LARGE\faEnvelope}\\[2pt]
     Classical\\message $y_1$};

\node[msgdesc, xshift=6pt, yshift=-1pt]
    (msg2) at (msgpos2)
    {{\LARGE\faEnvelope}\\[2pt]
     Classical\\message $y_2$};

\node[msgdesc, xshift=6pt, yshift=-1pt]
    (msgn) at (msgposn)
    {{\LARGE\faEnvelope}\\[2pt]
     Classical\\message $y_n$};

\coordinate (topstep1) at ([yshift=5pt]enc1.north);
\coordinate (topstep2) at ([yshift=5pt]enc2.north);
\coordinate (topstepn) at ([yshift=5pt]encn.north);
\coordinate (topend1) at (serverbox.west |- topstep1);
\coordinate (topend2) at (serverbox.west |- topstep2);
\coordinate (topendn) at (serverbox.west |- topstepn);
\draw[entlink] (enc1.north) -- (topstep1) -- (topend1);
\draw[entlink] (enc2.north) -- (topstep2) -- (topend2);
\draw[entlink] (encn.north) -- (topstepn) -- (topendn);
\node[entatom] (atom1)
    at (msg1.center |- topstep1) {\LARGE\faAtom};

\node[entatom] (atom2)
    at (msg2.center |- topstep2) {\LARGE\faAtom};

\node[entatom] (atomn)
    at (msgn.center |- topstepn) {\LARGE\faAtom};

\node[font=\large, align=center, anchor=south]
    at ([yshift=10pt]atom1.center)
    {Pre-shared\\entanglement};

\coordinate (outputport) at (serverbox.east |- decision);
\draw[arrow] (outputport) -- (decision.west);
\end{tikzpicture}}}
    \caption{Privacy-preserving hypothesis testing based on EACC.}
    \label{fig:eacc}
\end{figure}
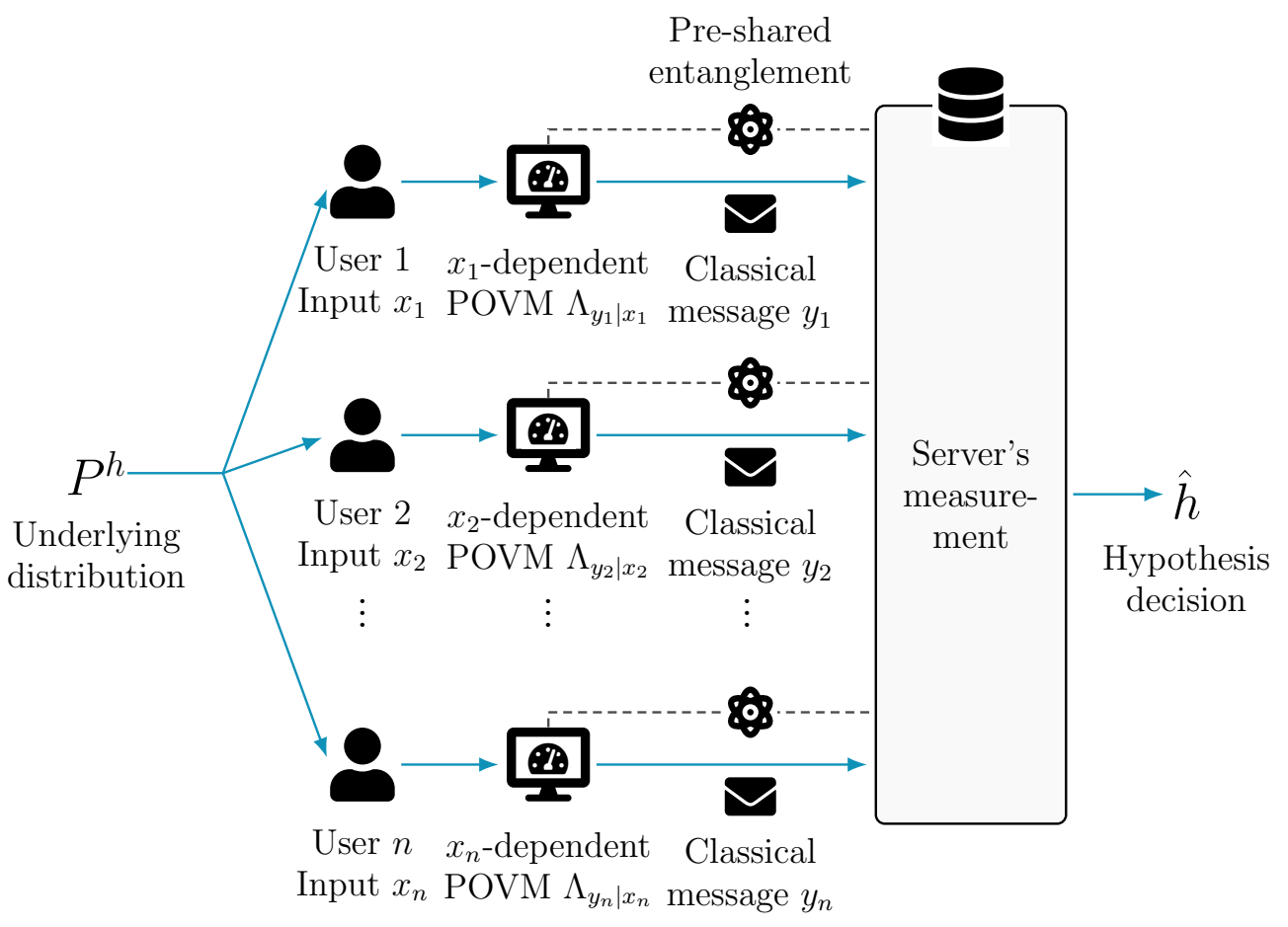

In the EACC-based scenario, depicted in Figure~\ref{fig:eacc}, each user $i$ and the server pre-share a bipartite state $\sigma_{A_iS_i}\in\mathcal D(\mathcal H_{A_i}\otimes\mathcal H_{S_i})$.
We assume that every user shares an independent copy of the same entangled state with the server, meaning that the overall entangled state shared between all users and the server is given by
\begin{equation}
    \sigma_{A^nS^n}:=\bigotimes_{i=1}^n\sigma_{A_iS_i},
\end{equation}
where $A^n:=A_1A_2\cdots A_n$ and $S^n := S_1\cdots S_n$.
Each $\sigma_{A_iS_i}$ is an identical copy of a fixed state $\sigma_{AS}$.
Conditioned on $X_i=x_i$, each user $i$ measures their local system $A_i$ using an input-dependent POVM $\{\Lambda_{y_i|x_i}\}_{y_i\in\mathcal Y}$ and sends the measurement result $Y_i\in\mathcal Y$ to the server. Every user uses the same family of input-dependent POVMs, and $\mathcal{Y}$ denotes a finite message alphabet.

This EACC protocol can be characterized by the pair 
\begin{equation}
    \Gamma:=\left(\sigma_{AS},\{\Lambda_{y|x}\}_{x\in\mathcal X,y\in\mathcal Y}\right),
\end{equation}
where $\sigma_{AS}\in\mathcal D(\mathcal H_A\otimes\mathcal H_S)$ and $\{\Lambda_{y|x}\}_{y\in\mathcal Y}$ is a POVM on system $A$ for each $x\in\mathcal X$.
If $X^n =x^n$, the server gets the classical-quantum state $\otimes_{i=1}^n \widehat Q_{x_i}(\Gamma)$, where
\begin{equation}\label{eq:eacc_induced_cq}
    \widehat Q_x(\Gamma):=\sum_{y\in\mathcal Y}\ketbra{y}\otimes\rho_{y|x}(\Gamma),
\end{equation}
with
\begin{equation}\label{eq:rho_yx}
    \rho_{y|x}(\Gamma):=\Tr_A[(\Lambda_{y|x}\otimes I_S)\sigma_{AS}].
\end{equation}

Under $P^{\eta, h}$ for $h \in [v]$, since the server does not have access to $X^n$, we can assume, without loss of generality, that the server gets the classical-quantum state $[\omega_{\eta, h}^{\mathrm{EACC}}(\Gamma)]^{\otimes n}$, where 
\begin{align}\label{eq:omega_eacc}
    \omega_{\eta, h}^{\mathrm{EACC}}(\Gamma) &:= \sum_{x\in[v]}P^{\eta, h}(x)\widehat Q_x(\Gamma)
    \\ &= \sum_{y\in\mathcal Y}\ketbra{y}\otimes\varphi_{y|\eta, h}(\Gamma),
\end{align}
with
\begin{equation}
    \varphi_{y|\eta, h}(\Gamma):=\sum_{x\in[v]}P^{\eta, h}(x)\rho_{y|x}(\Gamma).
\end{equation}
Under $P^0$, the corresponding state is $[\omega_0^{\mathrm{EACC}}(\Gamma)]^{\otimes n}$, where $\omega_0^{\mathrm{EACC}}(\Gamma)$ and $\varphi_{y|0}(\Gamma)$ are analogously defined by replacing $P^{\eta,h}$ with $P^0$ in the above expressions.
For brevity, we write $\omega_{\eta,h}(\Gamma):=\omega_{\eta,h}^{\mathrm{EACC}}(\Gamma)$ and $\omega_0(\Gamma):=\omega_0^{\mathrm{EACC}}(\Gamma)$.
The server performs a measurement on the received state to infer the underlying hypothesis.
The error exponents achievable by the EACC protocol $\Gamma$ are given by Eqs.~\eqref{eq:asym_stein} and~\eqref{eq:sym_Chernoff}.
To explicitly show the dependence of these error exponents on the EACC protocol, we write
\begin{align}
    A_\eta (\Gamma) &:= A(\{\omega_{\eta, h}(\Gamma)\}_{h \in [v]}, \omega_0(\Gamma)), \label{eq:EACC_exp_asym}
    \\ S_\eta (\Gamma) &:= S(\{\omega_{\eta, h}(\Gamma)\}_{h \in [v]}). \label{eq:EACC_exp_sym}
\end{align}

\subsection{Measurement Local Differential Privacy
and Privacy--Utility Tradeoffs}\label{sec:mldp}
Now, we propose a privacy constraint on EACC protocols which is analogous to the QLDP constraint on CQ channels.
One may try to impose the LDP constraint only on the likelihood of the transmitted message $Y$ given the input $X$.
However, protecting only $Y$ cannot sufficiently protect the private input $X$. 
This is because two distinct inputs can induce almost identical likelihoods while producing orthogonal post-measurement states on the server's system.
In this case, the server can extract additional information about the input by measuring the post-measurement state conditioned on the received message $Y$.
Hence, we require a privacy constraint defined over both the classical message and the post-measurement state.
Furthermore, even under the assumption that the measurements $\{\Lambda_{y|x}\}_{x\in\mathcal X,y\in\mathcal Y}$ can be ideally implemented by the users themselves, the shared entangled state might be prepared in an adversarial manner by the server. 
Therefore, the privacy constraint should encompass all possible bipartite states without making any assumptions about the server side.

Based on the above discussion, we consider an arbitrary finite-dimensional system $E$ on the server side, an arbitrary bipartite state $\rho_{AE}\in\mathcal D(\mathcal H_A\otimes\mathcal H_E)$, and a POVM $\{M_z\}_{z\in\mathcal Z}$ on $E$.
When the private input is $x$, the user applies $\{\Lambda_{y|x}\}_{y\in\mathcal Y}$ to $A$, and the resulting joint distribution of the message $y$ and the measurement outcome $z$ is given by
\begin{equation}
    P_{\rho,M}(y,z|x):=\Tr[(\Lambda_{y|x}\otimes M_z)\rho_{AE}].
\end{equation}
We formalize the privacy constraint on an EACC protocol using this conditional probability as follows.

\begin{definition}[Measurement local differential privacy]\label{def:mldp}
A family of POVMs $\{\{\Lambda_{y|x}\}_{y\in\mathcal Y}:x\in\mathcal X\}$ satisfies $\epsilon$-measurement local differential privacy ($\epsilon$-MLDP) if
\begin{equation}
    P_{\rho,M}(y,z|x)\le e^\epsilon P_{\rho,M}(y,z|x')
\end{equation}
for every finite-dimensional auxiliary system $E$, every state $\rho_{AE}\in\mathcal D(\mathcal H_A\otimes\mathcal H_E)$, every POVM $\{M_z\}_{z\in\mathcal Z}$ on $E$, and all $x,x'\in\mathcal X$, $y\in\mathcal Y$, and $z\in\mathcal Z$.
\end{definition}
We say that an EACC protocol $\Gamma=(\sigma_{AS},\{\Lambda_{y|x}\})$ is $\epsilon$-MLDP if its POVM family satisfies the $\epsilon$-MLDP condition, or simply refer to it as an $\epsilon$-MLDP EACC protocol.
The operational condition above admits the following simple, equivalent characterization, whose proof is in Appendix~\ref{app:mldp_operator}.
\begin{proposition}\label{prop:simplified_mldp}
A POVM family $\{\Lambda_{y|x}\}$ satisfies $\epsilon$-MLDP if and only if
\begin{equation}
    \Lambda_{y|x}\le e^\epsilon\Lambda_{y|x'},\qquad \forall x,x'\in\mathcal X,\ y\in\mathcal Y.
\end{equation}
\end{proposition}

To relate MLDP to the QLDP framework of Sec.~\ref{subsec:hypothesis_testing}, recall that an EACC protocol $\Gamma$ induces the CQ channel described in Eq.~\eqref{eq:eacc_induced_cq}.
If $\Gamma$ is $\epsilon$-MLDP, Proposition~\ref{prop:simplified_mldp} implies that
\begin{equation}
    \widehat Q_x(\Gamma)\le e^\epsilon\widehat Q_{x'}(\Gamma),\qquad \forall x,x'\in\mathcal X.
\end{equation}
Thus, every $\epsilon$-MLDP EACC protocol induces an $\epsilon$-QLDP channel.
The converse, however, does not necessarily hold; the QLDP constraint on $\widehat Q(\Gamma)$ for a fixed entangled state specified in $\Gamma$ may not hold for other entangled states.

We denote by
$\mathfrak M_\epsilon^{\mathrm{EACC}}(k,d)$
the set of all $\epsilon$-MLDP EACC protocols
$\Gamma=(\sigma_{AS},
\{\Lambda_{y|x}\}_{x\in\mathcal X,y\in\mathcal Y})$
with $\dim\mathcal H_A=\dim\mathcal H_S=d$
and $|\mathcal Y|=k$.
We also define
\begin{equation}
    \mathfrak M_\epsilon^{\mathrm{EACC}}
    :=
    \bigcup_{k,d\in\mathbb N}
    \mathfrak M_\epsilon^{\mathrm{EACC}}(k,d).
\end{equation}

For fixed $\eta\in(0,1]$, the PUTs under
the MLDP constraint are defined by
\begin{align}
    A_\eta[\mathfrak M_\epsilon^{\mathrm{EACC}}]
    &:=
    \sup_{\Gamma\in
    \mathfrak M_\epsilon^{\mathrm{EACC}}}
    A_\eta(\Gamma),\label{PUT_A}\\
    S_\eta[\mathfrak M_\epsilon^{\mathrm{EACC}}]
    &:=
    \sup_{\Gamma\in
    \mathfrak M_\epsilon^{\mathrm{EACC}}}
    S_\eta(\Gamma). \label{PUT_S}
\end{align}
For fixed $k$ and $d$, the corresponding PUTs are
defined by restricting the suprema in Eqs.~\eqref{PUT_A}
and \eqref{PUT_S} to $\mathfrak M_\epsilon^{\mathrm{EACC}}(k,d)$.

\subsection{MLDP EACC Protocols with Shared Randomness}\label{sec:shared_randomness}

Beyond sharing an entangled state between each user and the server, we consider an EACC protocol that uses pre-shared classical randomness.
While this classical randomness can be subsumed under the entangled state, we distinguish the two because classical resources are more cost-effective.
Additional classical randomness enables the users and server to agree beforehand on a specific EACC protocol, conditioned on the outcome of the randomness.

Formally, let $G$ be a random variable taking values in a finite set $\mathcal G$, with distribution $p_{G}$ independent of the private input.
An EACC protocol with shared randomness is denoted by
\begin{equation}
    \Gamma^{\mathrm{SR}}:=\left(p_{G},\{\Gamma_g\}_{g\in\mathcal G}\right),
\end{equation}
where $\Gamma_g$ is an EACC protocol for each $g \in \mathcal G$.

Each user--server pair independently draws a copy of $G$ according to $p_G$, and its realization is known to both parties.
The server gets a classical-quantum state $[\omega_{\eta, h}^{\mathrm{SR}}(\Gamma^{\mathrm{SR}})]^{\otimes n}$ under $P^{\eta,h}$ for $h\in[v]$, or $[\omega_{0}^{\mathrm{SR}} (\Gamma^{\mathrm{SR}})]^{\otimes n}$ under $P^0$, where
\begin{align}\label{eq:shared_randomness_state}
    \omega_{\eta, h}^{\mathrm{SR}}(\Gamma^{\mathrm{SR}})&:=\sum_{g\in\mathcal G}p_{G}(g)\ketbra{g}\otimes\omega_{\eta, h}^{\mathrm{EACC}}(\Gamma_g), \\
    \omega_0^{\mathrm{SR}}(\Gamma^{\mathrm{SR}})&:=\sum_{g\in\mathcal G}p_G(g)\ketbra{g}\otimes\omega_0^{\mathrm{EACC}}(\Gamma_g).
\end{align}

The privacy constraint on $\Gamma^{\mathrm{SR}}$ is defined in a manner analogous to Definition~\ref{def:mldp}, by considering shared classical randomness as a component of the entangled state.
By Proposition~\ref{prop:simplified_mldp}, it can be checked that $\Gamma^{\mathrm{SR}}$ is $\epsilon$-MLDP if and only if $\Gamma_g$ is $\epsilon$-MLDP for all $g \in \mathcal{G}$. 
Let
$\mathfrak M_{\epsilon,\mathrm{SR}}^{\mathrm{EACC}}(k,d)$
denote the set of all $\epsilon$-MLDP EACC protocols with shared randomness
$\Gamma^{\mathrm{SR}}
=(p_G,\{\Gamma_g\}_{g\in\mathcal G})$,
where
$\Gamma_g\in
\mathfrak M_\epsilon^{\mathrm{EACC}}(k,d)$
for every $g\in\mathcal G$.

The error exponents achievable by $\Gamma^{\mathrm{SR}}$ are  given by Eqs.~\eqref{eq:EACC_exp_asym} and~\eqref{eq:EACC_exp_sym}, with $\Gamma = \Gamma^{\mathrm{SR}}$ and the corresponding states $\omega_{\eta, h}^{\mathrm{SR}}(\Gamma^{\mathrm{SR}})$ and $\omega_0^{\mathrm{SR}}(\Gamma^{\mathrm{SR}})$.
The PUTs over
$\mathfrak M_{\epsilon,\mathrm{SR}}^{\mathrm{EACC}}(k,d)$
and $\mathfrak M_{\epsilon,\mathrm{SR}}^{\mathrm{EACC}}:=\bigcup_{k,d\in\mathbb N}
\mathfrak M_{\epsilon,\mathrm{SR}}^{\mathrm{EACC}}(k,d)$
are defined analogously to those over
$\mathfrak M_\epsilon^{\mathrm{EACC}}(k,d)$
and $\mathfrak M_\epsilon^{\mathrm{EACC}}$,
respectively.

When the local dimension is unrestricted,
shared randomness does not enlarge the
achievable PUTs. Indeed, any such protocol
can be represented as an ordinary MLDP
EACC protocol by incorporating the shared
random variable into orthogonal classical
registers of the bipartite state.
Consequently, the unrestricted PUTs
defined in Sec.~\ref{sec:mldp} remain unchanged
when shared randomness is allowed.
However, this argument does not establish
the same equivalence for fixed $d$,
since incorporating the classical registers
may increase the local dimension.

\section{Summary of Main Results}
\label{sec:summary}

We first establish that any QLDP CQ channel can be
simulated by an MLDP EACC protocol without loss
in the error exponents for either hypothesis-testing
task.

\begin{theorem}\label{thm:simulation}
    For every $\epsilon>0$ and every
    $\epsilon$-QLDP channel
    $Q\in\mathfrak M_\epsilon^{\mathrm{CQ}}(d)$,
    there exists an $\epsilon$-MLDP EACC protocol
    $\Gamma\in
    \mathfrak M_\epsilon^{\mathrm{EACC}}(d^2,d)$
    such that, for all $\eta\in(0,1]$,
    \begin{equation}
        A_\eta(Q)=A_\eta(\Gamma),
        \qquad
        S_\eta(Q)=S_\eta(\Gamma).
    \end{equation}
\end{theorem}

Theorem~\ref{thm:simulation} is established by
constructing an MLDP EACC simulation of an
arbitrary QLDP CQ channel using generalized
quantum teleportation~\cite{gen_quan_telep}.
The construction preserves both error exponents
while satisfying the MLDP constraint.
The details are given in Sec.~\ref{sec:eacc_sim}.

As shown in Sec.~\ref{sec:mldp}, every MLDP EACC protocol
induces a QLDP CQ channel, implying that the EACC
PUTs cannot exceed the corresponding CQ PUTs.
Conversely, Theorem~\ref{thm:simulation} guarantees
that every QLDP CQ channel admits an MLDP EACC
simulation preserving both error exponents.
Therefore, the two frameworks achieve identical
PUTs when communication and entanglement resources
are unrestricted. Moreover, the optimal EACC PUTs can be
asymptotically achieved as $\epsilon\to0$
with finite classical communication and
entanglement resources, as stated below.

\begin{corollary}\label{cor:cq_eacc_equiv}
    For all $v\ge2$, $\epsilon>0$, and
    $\eta\in(0,1]$, we have
    \begin{align}
        A_\eta[\mathfrak M_\epsilon^{\mathrm{EACC}}]
        &=
        A_\eta[\mathfrak M_\epsilon^{\mathrm{CQ}}],\\
        S_\eta[\mathfrak M_\epsilon^{\mathrm{EACC}}]
        &=
        S_\eta[\mathfrak M_\epsilon^{\mathrm{CQ}}].
    \end{align}
    Consequently, for $v\ge3$,
    \begin{align}
        \lim_{\epsilon\to0}
        \frac{
            A_\eta[\mathfrak M_\epsilon^{\mathrm{EACC}}]
        }{
            A_\eta[\mathfrak M_\epsilon^{\mathrm C}]
        }
        &=
        \lim_{\epsilon\to0}
        \frac{
            S_\eta[\mathfrak M_\epsilon^{\mathrm{EACC}}]
        }{
            S_\eta[\mathfrak M_\epsilon^{\mathrm C}]
        }
        \nonumber\\
        &=
        \frac{v(v-1)}
        {2\lfloor v/2\rfloor\lceil v/2\rceil}
        \ge\frac32.
    \end{align}
   Moreover, the optimal EACC PUTs can be
asymptotically achieved as $\epsilon\to0$
with a classical message alphabet of size
$d^2$ and local dimension $d$ satisfying
\begin{equation}
    d\ge
    2^{\max\{1,\lceil(v-2)/2\rceil\}}.
\end{equation}
Specifically, for any such $d$,
\begin{align}
    \lim_{\epsilon\to0}
    \frac{
        A_\eta[
            \mathfrak M_\epsilon^{\mathrm{EACC}}(d^2,d)
        ]
    }{
        A_\eta[
            \mathfrak M_\epsilon^{\mathrm{EACC}}
        ]
    }
    &=1,\\
    \lim_{\epsilon\to0}
    \frac{
        S_\eta[
            \mathfrak M_\epsilon^{\mathrm{EACC}}(d^2,d)
        ]
    }{
        S_\eta[
            \mathfrak M_\epsilon^{\mathrm{EACC}}
        ]
    }
    &=1.
\end{align}
\end{corollary}

The universal simulation in
Theorem~\ref{thm:simulation} uses a classical
message alphabet of size $d^2$.
We next show that the asymptotically optimal
PUTs can instead be achieved with only two
possible classical messages. 

\begin{theorem}\label{thm:simulation_onebit}
    For fixed $v\ge3$, $\eta\in(0,1]$,
    and any integer $d$ satisfying $d\ge 2^{\lceil(v-1)/2\rceil}$,
    \begin{align}
        \lim_{\epsilon\to0}
        \frac{A_\eta[\mathfrak M_\epsilon^{\mathrm{EACC}}(2,d)]}
        {A_\eta[
            \mathfrak M_\epsilon^{\mathrm{EACC}}
        ]}
        &=1,\\
        \lim_{\epsilon\to0}
        \frac{S_\eta[\mathfrak M_\epsilon^{\mathrm{EACC}}(2,d)]}
        {S_\eta[
            \mathfrak M_\epsilon^{\mathrm{EACC}}
        ]}
        &=1.
    \end{align}
\end{theorem}

Theorem~\ref{thm:simulation_onebit} shows that one bit of classical communication per user suffices to achieve the optimal PUTs in the high-privacy regime.
To establish this result, 
we exploit an additional symmetry of the
isoclinic mechanisms constructed from
Radon--Hurwitz EITFFs
~\cite[Thm.~4.1]{fickus2025}.
This symmetry enables exact EACC simulation
using only one bit of classical communication,
while retaining the asymptotically optimal
error exponents of the QLDP mechanisms.
The required local dimension, however,
can grow exponentially with the input
alphabet size $v$.
The construction is presented in
Sec.~\ref{sec:eacc_sim_opt}.

We finally investigate whether a strict quantum
advantage persists under simultaneous constraints
on entanglement and classical communication:
each user shares only a single entangled qubit pair
with the server and transmits one classical bit,
with shared randomness allowed when needed.
The following theorem establishes such an
advantage for both hypothesis-testing tasks
and every nonbinary input alphabet in the
high-privacy regime.

\begin{theorem}\label{thm:qubit_advantage}
For any fixed $v\ge 3$ and $\eta\in(0,1]$,
\begin{align}
\liminf_{\epsilon\to0}
\frac{A_\eta[\mathfrak M_\epsilon^{\mathrm{EACC}}(2,2)]}
     {A_\eta[\mathfrak M_\epsilon^{\mathrm C}]}
&\ge f(v),\\
\liminf_{\epsilon\to0}
\frac{S_\eta[\mathfrak M_{\epsilon,\mathrm{SR}}^{\mathrm{EACC}}(2,2)]}
     {S_\eta[\mathfrak M_\epsilon^{\mathrm C}]}
&\ge f(v),
\end{align}
where, with $m=\lfloor v/4\rfloor$,
\begin{equation}\label{eq:f}
f(v)=
\begin{cases}
\dfrac32,
& v=4m,\\[4pt]
\dfrac{3(4m+1)}{4(2m+1)},
& v=4m+1,\\[4pt]
\dfrac{3m+1}{2m+1},
& v=4m+2,\\[4pt]
\dfrac{3(m+1)(4m+3)}
      {2(2m+1)(2m+3)},
& v=4m+3.
\end{cases}
\end{equation}
Moreover, $f(v)\ge 5/4$ for all $v\ge3$, and
$\lim_{v\to\infty}f(v)=3/2$.
\end{theorem}

Theorem~\ref{thm:qubit_advantage} is proved
by constructing MLDP EACC protocols based
on the geometry of qubit measurements.
We arrange the corresponding Bloch vectors
in tetrahedral configurations to control
the privacy constraint while obtaining
favorable error exponents.
For symmetric hypothesis testing,
shared randomness is used to construct
the required protocol.
The resulting exponents strictly exceed
the optimal classical LDP exponents for
every $v\ge3$, demonstrating that quantum
advantage persists even under simultaneous
one-bit communication and single-qubit-pair
entanglement constraints.
The proof is given in Sec.~\ref{sec:qubit-onebit mech}.

\section{EACC Protocol Constructions and Analysis}\label{sec:main}

In this section, we present the constructions and
proofs underlying the main results of
Sec.~\ref{sec:summary}.
In Sec.~\ref{sec:eacc_sim}, we establish the exact
simulation of QLDP CQ channels by MLDP EACC
protocols, proving Theorem~\ref{thm:simulation},
and derive the unrestricted equivalence of CQ
and EACC PUTs stated in
Corollary~\ref{cor:cq_eacc_equiv}. 
In Sec.~\ref{sec:eacc_sim_opt}, we prove
Theorem~\ref{thm:simulation_onebit} by constructing
one-bit EACC simulations of asymptotically optimal
QLDP mechanisms.
Finally, in Sec.~\ref{sec:qubit-onebit mech}, we develop
EACC protocols using a single entangled qubit pair
and one bit of classical communication per user,
analyze their error exponents, and incorporate
shared randomness for symmetric hypothesis testing.
We then construct weighted tetrahedral Bloch-vector
constellations and compare the resulting error
exponents with the corresponding classical PUTs,
thereby completing the proof of
Theorem~\ref{thm:qubit_advantage}.

\subsection{Exact EACC Simulation of QLDP CQ Channels}
\label{sec:eacc_sim}

We first study when an EACC protocol can reproduce a CQ channel without transmitting the quantum state itself.

\subsubsection{A General Criterion for Exact Channel Simulation}

Let $Q:\mathcal X\to\mathcal D(\mathcal H_B)$ be a CQ channel.
The output of an EACC protocol and that of the CQ channel are not directly comparable because they live on different systems:
the EACC output consists of a classical message together with the server's quantum system, whereas the CQ channel outputs a quantum state on $B$.
We therefore introduce a server-side decoding map that converts the EACC output into a state on the CQ channel's output system $B$.

We say that an EACC protocol $\Gamma$ with $|\mathcal Y|=k$ simulates $Q$ exactly if there exists an input-independent completely positive trace-preserving (CPTP) map
\begin{equation}
    \mathcal T:
    \mathcal L(\mathbb C^k\otimes\mathcal H_S)
    \to
    \mathcal L(\mathcal H_B)
\end{equation}
such that
\begin{equation}\label{eq:exact_simulation}
    Q=\mathcal T\circ\widehat Q(\Gamma).
\end{equation}
We will call $\mathcal T$ a decoding map.

Our approach to exact CQ-channel simulation via an EACC protocol is based on a standard remote-state-preparation principle~\cite[Thm.~1]{mullan2012}.
The following lemma gives a sufficient condition for such an exact simulation while preserving local privacy.

\begin{lemma}\label{lem:unitary_average_simulation}
Let $Q=(Q_x)_{x\in\mathcal X}$ be a CQ channel on a $d$-dimensional system $B$.
Suppose that, for some $\tau\in\mathcal D(\mathcal H_B)$, there exist $k$ unitaries
$\{U_y\}_{y\in\mathcal Y}$, with $|\mathcal Y|=k$, such that
\begin{equation}\label{eq:unitary_average}
    \frac1k\sum_{y\in\mathcal Y}U_y^\dagger Q_xU_y=\tau,
    \qquad
    \forall x\in\mathcal X.
\end{equation}
Then there exists a protocol $\Gamma$ with a classical message alphabet of size $k$ and local dimension $d$ that simulates $Q$ exactly.
Moreover, if $Q$ is $\epsilon$-QLDP, $\Gamma$ can be chosen to be $\epsilon$-MLDP.
\end{lemma}
A sketch of the proof is as follows.
The protocol uses a purification of $\tau$ as the shared state. 
Conditioned on message $y$, the server's post-measurement state is $U_y^\dagger Q_xU_y$, so the server recovers $Q_x$ by applying $U_y$. Moreover, if $Q$ is $\epsilon$-QLDP, the QLDP operator inequality guarantees MLDP for this protocol.
The detailed proof is given in Appendix~\ref{app:proof_unitary_average_simulation}.

Generalized teleportation provides a universal instance of Lemma~\ref{lem:unitary_average_simulation} that applies to every finite-dimensional CQ channel.

\begin{example}[Generalized teleportation]
\label{ex:generalized_teleportation}
Let
$\{W_{ab}:a,b\in\{0,\ldots,d-1\}\}$
be the Heisenberg--Weyl operators defined by
\begin{equation}
    W_{ab}\ket{j}:=e^{2\pi iaj/d}\ket{j+b},
\end{equation}
where addition is modulo $d$.
They satisfy
\begin{equation}\label{eq:hw_average}
    \frac1{d^2}\sum_{a,b}W_{ab}^* Q_xW_{ab}^T=\frac1dI_d,
    \qquad
    \forall x\in\mathcal X.
\end{equation}
With $U_{ab}:=W_{ab}^T$, Eq.~\eqref{eq:unitary_average} holds with
$k=d^2$ and $\tau=I_d/d$.
Lemma~\ref{lem:unitary_average_simulation} therefore gives an exact EACC simulation of every $d$-dimensional CQ channel using a classical message alphabet of size $d^2$.
\end{example}

For the protocol $\Gamma$ constructed in Example~\ref{ex:generalized_teleportation},
Lemma~\ref{lem:unitary_average_simulation} guarantees the existence of a decoding map $\mathcal T$ such that $Q=\mathcal T\circ\widehat Q(\Gamma)$.
Moreover, for this construction, there exists an input-independent CPTP map $\mathcal E$ satisfying $\widehat Q(\Gamma)=\mathcal E\circ Q$.
The detailed protocol and explicit constructions of both maps are provided in Appendix~\ref{app:eff_gen_quant_tel}.

By linearity, the corresponding hypothesis-dependent states satisfy
\begin{align}
    \omega_{\eta, h}^{\mathrm{EACC}}(\Gamma)
    &=\mathcal E\bigl(\omega_{\eta, h}^{\mathrm{CQ}}(Q)\bigr),\\
    \omega_{\eta, h}^{\mathrm{CQ}}(Q)
    &=\mathcal T\bigl(\omega_{\eta, h}^{\mathrm{EACC}}(\Gamma)\bigr)
\end{align}
for every $h \in [v]$. 
The same relations hold for the uniform alternative.
Applying the data-processing inequalities for the quantum relative entropy and the quantum Chernoff information~\cite{Tomamichel2016} and using Eqs.~\eqref{eq:asym_stein} and~\eqref{eq:sym_Chernoff}, we obtain
\begin{align}
    A_\eta (Q)&\le A_\eta (\Gamma)\le A_\eta (Q),\\
    S_\eta (Q)&\le S_\eta (\Gamma)\le S_\eta (Q).
\end{align}
Therefore, $A_\eta (Q)=A_\eta (\Gamma)$ and $S_\eta (Q)=S_\eta (\Gamma)$.
When $Q$ is $\epsilon$-QLDP, Lemma~\ref{lem:unitary_average_simulation} also guarantees that the constructed protocol is $\epsilon$-MLDP. 
This proves Theorem~\ref{thm:simulation}.

\subsubsection{Equivalence of CQ and EACC PUTs}

Without resource constraints, EACC achieves the same PUTs as CQ mechanisms.
We state this equivalence for any utility that cannot increase under server-side quantum post-processing.

\begin{proposition}[CQ--EACC equivalence]
\label{prop:cq_eacc_equivalence}
Let $\mathcal U$ be a utility functional on finite-dimensional CQ channels satisfying
\begin{equation}\label{eq:utility_dpi}
    \mathcal U(\mathcal T\circ Q)
    \le
    \mathcal U(Q)
\end{equation}
for every CQ channel $Q$ and every CPTP map $\mathcal T$ acting on its output.
Then, for every $\epsilon>0$,
\begin{equation}\label{eq:general_cq_eacc_equivalence}
    \sup_{\Gamma\in\mathfrak M_\epsilon^{\mathrm{EACC}}}
    \mathcal U(\widehat Q(\Gamma))
    =
    \sup_{Q\in\mathfrak M_\epsilon^{\mathrm{CQ}}}
    \mathcal U(Q).
\end{equation}
\end{proposition}

\begin{proof}
Every $\epsilon$-MLDP EACC protocol $\Gamma$ induces an $\epsilon$-QLDP CQ channel $\widehat Q(\Gamma)$, as shown in Sec.~\ref{sec:mldp}.
Hence,
\begin{equation}
    \sup_{\Gamma\in\mathfrak M_\epsilon^{\mathrm{EACC}}}
    \mathcal U(\widehat Q(\Gamma))
    \le
    \sup_{Q\in\mathfrak M_\epsilon^{\mathrm{CQ}}}
    \mathcal U(Q).
\end{equation}

Conversely, let $Q$ be any finite-dimensional $\epsilon$-QLDP CQ channel.
Example~\ref{ex:generalized_teleportation} and Lemma~\ref{lem:unitary_average_simulation} give an $\epsilon$-MLDP EACC protocol $\Gamma_Q$ and an input-independent CPTP map $\mathcal T$ such that
\begin{equation}
    Q=\mathcal T\circ\widehat Q(\Gamma_Q).
\end{equation}
By Eq.~\eqref{eq:utility_dpi},
\begin{equation}
    \mathcal U(Q)
    \le
    \mathcal U(\widehat Q(\Gamma_Q)).
\end{equation}
Taking the supremum over $Q$ gives the reverse inequality.
\end{proof}

Since both error exponents satisfy Eq.~\eqref{eq:utility_dpi}~\cite{Tomamichel2016}, Proposition~\ref{prop:cq_eacc_equivalence} shows that EACC causes no loss in the PUTs when resources are unrestricted.
Together with Theorem~\ref{thm:known_optima}, this proves Corollary~\ref{cor:cq_eacc_equiv}.

The remaining question is what classical and quantum resources are needed to retain these PUTs.
We next show that a class of optimal QLDP mechanisms in the high-privacy regime can instead be implemented by EACC protocols using only two possible classical messages, i.e., one classical bit.

\subsection{One-Bit Exact EACC Simulation of Asymptotically Optimal QLDP Mechanisms}
\label{sec:eacc_sim_opt}

A reduction in classical communication is possible when the input-independent average in Eq.~\eqref{eq:unitary_average} can be obtained using fewer unitaries.
To prove Theorem~\ref{thm:simulation_onebit}, we exploit this observation by constructing an isoclinic mechanism from a Radon--Hurwitz EITFF whose projections can all be mapped to their complements by the same unitary.

Under the stronger condition $v\le N_{\mathrm{RH}}(r)+1$, which can be satisfied by choosing $r=2^{\lceil(v-3)/2\rceil}$, the Radon--Hurwitz construction provides an $\mathrm{EITFF}(2r,r,v)$ in the following block form~\cite[Thms.~3.1, 4.1(b)]{fickus2025}.
Specifically, after a suitable choice of basis, for $x\in[v-1]$,
\begin{equation}\label{eq:rh_Fx}
    F_x=
    \begin{pmatrix}
        \cos^2\theta I_r
        &
        -\sin\theta\cos\theta J_x\\
        \sin\theta\cos\theta J_x
        &
        \sin^2\theta I_r
    \end{pmatrix},
\end{equation}
where
\begin{equation}
    \cos^2\theta
    =
    \frac{v-2}{2(v-1)},
\end{equation}
and each $J_x$ is a skew-Hermitian unitary, while
\begin{equation}\label{eq:rh_Fv}
    F_v=
    \begin{pmatrix}
        I_r&0\\
        0&0
    \end{pmatrix}.
\end{equation}
These block forms ensure that the isoclinic mechanism based on Radon--Hurwitz EITFFs described above satisfies Eq.~\eqref{eq:unitary_average} with $k=2$ for every input.

\begin{corollary}[One-bit exact channel simulation]
\label{cor:k2_exact}
For $\epsilon>0$ and $v\ge3$, there exist an $\epsilon$-QLDP isoclinic mechanism $Q^{\mathrm{iso}}$ from the family above and a protocol
$\Gamma\in\mathfrak M_\epsilon^{\mathrm{EACC}}(2,2^{\lceil (v-1)/2 \rceil})$
that exactly simulates $Q^{\mathrm{iso}}$.
\end{corollary}
\begin{proof}
Choose $r = 2^{\lceil (v-3)/2 \rceil}$ and consider the Radon--Hurwitz $\mathrm{EITFF}(2r,r,v)$ described above.
Define
\begin{equation}
    U_1:=I_{2r},
    \qquad
    U_2:=
    \begin{pmatrix}
        0&I_r\\
        I_r&0
    \end{pmatrix}.
\end{equation}
From Eqs.~\eqref{eq:rh_Fx} and~\eqref{eq:rh_Fv}, we have
\begin{equation}
    U_2^\dagger F_xU_2
    =
    I_{2r}-F_x,
    \qquad
    \forall x\in[v].
\end{equation}
Therefore, using Eq.~\eqref{eq:isoclinic_mechanism},
\begin{align}
    \!\!\frac12\sum_{y=1}^2
    U_y^\dagger Q_x^{\mathrm{iso}}U_y
    &=
    \frac{\mu_{\mathrm{iso}}}{2r}I_{2r}
    +
    \frac{1-\mu_{\mathrm{iso}}}{2r}
    \left(F_x+U_2^\dagger F_xU_2\right)
    \nonumber\\
    &=
    \frac1{2r}I_{2r},
    \qquad
    \forall x\in[v].
\end{align}
Hence the condition in Eq.~\eqref{eq:unitary_average} holds with
\begin{equation}
    k=2,
    \qquad
    \tau=\frac1{2r}I_{2r}.
\end{equation}
Lemma~\ref{lem:unitary_average_simulation} therefore gives an
$\epsilon$-MLDP EACC protocol
\begin{equation}
    \Gamma\in
    \mathfrak M_\epsilon^{\mathrm{EACC}}(2,2r)
\end{equation}
that exactly simulates $Q^{\mathrm{iso}}$.
\end{proof}

Corollary~\ref{cor:k2_exact} shows that the error exponents achieved by the isoclinic mechanisms constructed above can be realized by EACC protocols using only one bit of classical communication.
Together with Corollary~\ref{cor:cq_eacc_equiv}, this implies that the PUTs under the $\epsilon$-MLDP constraint in the high-privacy regime can be achieved by EACC protocols using one-bit communication.
The construction extends to any larger local dimension by embedding the shared state and setting each POVM element to $I/2$ on the unused subspace.
This preserves MLDP and both error exponents, proving Theorem~\ref{thm:simulation_onebit}.
However, the local dimension required by this one-bit exact-simulation construction is $d=2^{\lceil(v-1)/2\rceil}$, which grows exponentially with the input alphabet size $v$.
This motivates the dimension-constrained EACC protocols studied next.

\subsection{Qubit EACC Protocols with One-Bit Communication}
\label{sec:qubit-onebit mech}
\subsubsection{Protocol Construction and Error Exponents}
Here, we investigate whether a quantum advantage survives when both the transmitted-message alphabet size and the local quantum dimension are reduced to their smallest nontrivial values.
We therefore fix $d=k=2$, so that each user shares one entangled qubit pair with the server and transmits one classical bit.
In this regime, exact channel simulation of the isoclinic mechanisms constructed from Radon--Hurwitz EITFFs is no longer available for arbitrary $v$.\footnote{For $d=2$, a Radon--Hurwitz EITFF reduces to an equiangular tight frame, and Gerzon's bound limits the existence of Radon--Hurwitz EITFFs to $v\le4$~\cite[Thm.~12.2]{waldron2018finite}.}
We therefore design EACC protocols with $d=k=2$ directly.

The basic idea is to encode each private input $x$ through a measurement direction on the Bloch sphere, adjusting how strongly the user's measurement depends on this direction.
The geometry of the Bloch vectors therefore determines both how strongly different inputs can be distinguished and how restrictive the privacy constraint is.

Let $\mathcal R:=(\mathbf r_x)_{x\in[v]}$ be a collection of unit Bloch vectors $\mathbf r_x \in \mathbb R^3$. We define the associated EACC protocol with local qubit systems and one-bit communication by
\begin{equation}
    \Gamma_{\mathcal R}
    :=\left(\Phi_{AS}^2,\{\Lambda_{y|x}^{\mathcal R}\}_{x\in[v],y\in[2]}\right),
\end{equation}
where
\begin{align}
    \Lambda_{y|x}^{\mathcal R}
    &:=\frac{1-\mu}{2}I_2+\mu P_{x,y},\\
    P_{x,y}
    &:=\frac12\left(I_2+(-1)^{y-1}\mathbf r_x\cdot\bm\sigma\right),
    \qquad y\in[2],
\end{align}
$\bm\sigma=(\sigma_1,\sigma_2,\sigma_3)$ is the vector of Pauli matrices, and $\mu\in(0,1]$ controls how strongly the POVM depends on the private input $x$. 
Because $P_{x,1}+P_{x,2}=I_2$, $\{\Lambda_{y|x}^{\mathcal R}\}_{y\in[2]}$ is a POVM for every $x$.

Using Eq.~\eqref{eq:rho_yx}, the subnormalized state associated with input $x$ and message $y$ is
\begin{equation}\label{eq:binary_steered_state}
    \rho_{y|x}(\Gamma_{\mathcal R})=\frac14I_2+\frac\mu4(-1)^{y-1}\tilde{\mathbf r}_x\cdot\bm\sigma,
\end{equation}
where
\begin{equation}
    \tilde {\mathbf r}_x:=(r_{x,1},-r_{x,2},r_{x,3}).
\end{equation}
Note that $\mathbf r_x \cdot \mathbf r_{x'} = \tilde {\mathbf r}_x \cdot \tilde {\mathbf r}_{x'}$ for any $x, x' \in [v]$. 

For distinct $x,x'\in [v]$, define
\begin{equation}\label{eq:pairwise_overlap}
    c_{x,x'} :=\Tr(P_{x,1}P_{x',1}) =\frac{1+\mathbf r_x\cdot\mathbf r_{x'}}2.
\end{equation}
We further set
\begin{equation}
    c_{\min}:=\min_{x\ne x'}c_{x,x'},
    \qquad
    c_{\max}:=\max_{x\ne x'}c_{x,x'}.
\end{equation}

The privacy constraint is determined by the pair of input-dependent measurements that are most distinguishable.
In the Bloch-sphere representation, this corresponds to the pair of vectors with the largest separation, or equivalently, the smallest overlap $c_{\min}$.
The following lemma makes this dependence explicit.

\begin{lemma}\label{lem:exact_privacy}
The protocol $\Gamma_{\mathcal R}$ satisfies $\epsilon$-MLDP if and only if
\begin{equation}\label{eq:mu_privacy_bound}
    \mu\le\frac{e^\epsilon-1} {\sqrt{(e^\epsilon-1)^2+4e^\epsilon(1-c_{\min})}}.
\end{equation}
\end{lemma}

The proof is given in Appendix~\ref{app:binary_privacy}.
For the remainder of this paper, we choose $\mu$ to saturate the privacy bound in Eq.~\eqref{eq:mu_privacy_bound}.
With this choice, $\Gamma_{\mathcal R}\in\mathfrak M_\epsilon^{\mathrm{EACC}}(2,2)$. 

To simplify the hypothesis-dependent states, we restrict attention to constellations for which the Bloch-vector contributions cancel under the uniform distribution.
Specifically, we call a constellation $\mathcal R=(\mathbf r_x)_{x\in[v]}$ balanced if
\begin{equation}\label{eq:balanced_constellation}
    \sum_{x=1}^v\mathbf r_x=0.
\end{equation}
Under the uniform alternative $P^0$,
\begin{equation}
    \varphi_{y|0}(\Gamma_{\mathcal R})=\frac14I_2.
\end{equation}
Under the biased hypothesis $P^{\eta,h}$ for $h\in[v]$,
\begin{equation}\label{eq:varphi_yh}
    \varphi_{y|\eta, h}(\Gamma_{\mathcal R})  =\frac14I_2+\frac{\mu\eta}{4}(-1)^{y-1}\tilde{\mathbf r}_h\cdot\bm\sigma.
\end{equation}
Then the state in Eq.~\eqref{eq:omega_eacc} is
\begin{equation}\label{eq:binary_server_state}
    \omega_{\eta, h}(\Gamma_{\mathcal R})=\sum_{y\in[2]}\ketbra{y}\otimes\varphi_{y|\eta, h}(\Gamma_{\mathcal R}).
\end{equation}

\begin{theorem}\label{thm:exact_asym}
For any fixed $\eta\in(0,1]$ and any balanced constellation $\mathcal R$, the error exponent of $\Gamma_{\mathcal R}$ for asymmetric hypothesis testing is
\begin{align}
    &A_\eta \left(\Gamma_{\mathcal R}\right)\nonumber\\
    &\quad=\frac{1+\mu\eta}{2}\log(1+\mu\eta)+\frac{1-\mu\eta}{2}\log(1-\mu\eta).
\end{align}
Moreover,
\begin{equation}
    \lim_{\epsilon\to0}\frac{A_\eta (\Gamma_{\mathcal R})}{\epsilon^2}=\frac{\eta^2}{8(1-c_{\min})}.
\end{equation}
\end{theorem}

The above exponent depends on the constellation only through $c_{\min}$. We next turn to symmetric hypothesis testing.

\begin{theorem}\label{thm:exact_sym}
For any fixed $\eta\in(0,1]$ and any balanced constellation $\mathcal R$, the error exponent of $\Gamma_{\mathcal R}$ for symmetric hypothesis testing is
\begin{equation}
    S_\eta \left(\Gamma_{\mathcal R}\right)
    =-\log\left[c_{\max}+(1-c_{\max})\sqrt{1-\mu^2\eta^2}\right].
\end{equation}
Moreover,
\begin{equation} \label{eq:deter_sym}
    \lim_{\epsilon\to0} \frac{S_\eta (\Gamma_{\mathcal R})}{\epsilon^2}=\frac{\eta^2(1-c_{\max})}{8(1-c_{\min})}.
\end{equation}
\end{theorem}

The proofs of Theorems~\ref{thm:exact_asym} and~\ref{thm:exact_sym} are given in Appendices~\ref{app:asym_proof} and~\ref{app:sym_proof}, respectively.
The appearance of $c_{\max}$ reflects the difference between the two testing tasks.
For the hypotheses considered here, asymmetric hypothesis testing compares each biased hypothesis only with the uniform alternative, whereas symmetric hypothesis testing must distinguish every pair of biased hypotheses.
The error exponent for symmetric hypothesis testing is therefore limited by the least distinguishable pair.

\subsubsection{Permutation Ensemble for Symmetric Hypothesis Testing}

For an input-to-Bloch-vector assignment that is fixed across users, $c_{\max}$ approaches one as the alphabet size grows.
Indeed,
\begin{equation}
    1-c_{\max}
    =\frac14\min_{x\ne x'}\|\mathbf r_x-\mathbf r_{x'}\|^2.
\end{equation}
As more input symbols are assigned to the Bloch sphere, some pair of vectors must become arbitrarily close; if two symbols share the same vector, then $c_{\max}=1$ and the error exponent for symmetric hypothesis testing is already zero.
To remove this limitation, we use the shared-randomness framework described in Sec.~\ref{sec:shared_randomness} to independently permute the assignment of Bloch vectors to input symbols for each user.

\begin{definition}[Permutation ensemble]\label{def:ensemble}
Let $S_v$ denote the permutation group on $[v]$, and let $G$ be uniformly distributed over $S_v$.
Given $G=\pi$ and a protocol $\Gamma = ( \sigma_{AS}, \{\Lambda_{y | x}\} )$, the user employs the protocol
$\Gamma_\pi:= ( \sigma_{AS}, \{\Lambda_{y | \pi(x)}\} )$.
This defines the protocol
$\Gamma^{\mathrm{perm}}:=(p_{G}, \{ \Gamma_\pi\}_{\pi \in S_v})$, where $p_{G}(\pi)=1/v!$ for every $\pi\in S_v$.
\end{definition}

When $\Gamma=\Gamma_{\mathcal R}$ is the protocol associated with a constellation $\mathcal R$, we write the resulting permutation-ensemble protocol as $\Gamma^{\mathrm{perm}}_{\mathcal R}$.
Each realized permutation is an input-independent relabeling of the original POVM family, so $\epsilon$-MLDP is preserved.
The statistical role of permutation is to prevent any fixed pair of hypotheses from being consistently associated with the same pair of Bloch vectors.
A uniformly random permutation maps a fixed pair $(h,h')$ with $h \ne h'$ uniformly over all ordered pairs of distinct input indices.

For a balanced constellation, $0=\|\sum_x\mathbf r_x\|^2=v+\sum_{x\ne x'}\mathbf r_x\cdot\mathbf r_{x'}$, so the resulting average pairwise overlap is
\begin{equation}\label{eq:average_overlap}
    \bar c
    :=\frac1{v(v-1)}\sum_{x\ne x'}c_{x,x'}
    =\frac{v-2}{2(v-1)}.
\end{equation}

Since the realized permutation is known to the server, each observation can still be interpreted using the corresponding assignment.
As shown below, under the permutation ensemble, the error exponent for symmetric hypothesis testing is governed by $\bar c$ rather than by $c_{\max}$.

\begin{theorem}\label{thm:exact_ens}
For any fixed $\eta\in(0,1]$ and any balanced constellation $\mathcal R$, the error exponent for symmetric hypothesis testing under the permutation ensemble is
\begin{equation}
    S_\eta\left(\Gamma_{\mathcal R}^{\mathrm{perm}}\right)
    =-\log\left[\bar c+(1-\bar c)\sqrt{1-\mu^2\eta^2}\right].
\end{equation}
Moreover,
\begin{equation}
    \lim_{\epsilon\to0} \frac{S_\eta (\Gamma_{\mathcal R}^{\mathrm{perm}})}{\epsilon^2} =\frac{v\eta^2}{16(v-1)(1-c_{\min})}.
\end{equation}
\end{theorem}

The proof is given in Appendix~\ref{app:ensemble_proof}.

\subsubsection{Geometric Design of Weighted Tetrahedral Constellations}

Our remaining task is to choose the Bloch vectors so as to maximize $c_{\min}$ while preserving the balance condition.
For balanced constellations, Theorems~\ref{thm:exact_asym} and~\ref{thm:exact_ens} show why this is the relevant geometric objective: a larger $c_{\min}$ allows a larger value of $\mu$ under the $\epsilon$-MLDP constraint, which in turn yields larger error exponents.
To guide this design, we first derive an upper bound on $c_{\min}$ and identify the cases in which the bound is attained.

\begin{lemma}\label{lem:cmin_bound}
For any collection $\mathcal R=(\mathbf r_x)_{x\in[v]}$ of $v\ge3$ unit vectors $\mathbf r_x\in\mathbb R^3$ satisfying $\sum_x\mathbf r_x=0$,
\begin{equation}
    c_{\min}\le
    \begin{cases}
        \frac14, & v=3,\\
        \frac13, & v\ge4.
    \end{cases}
\end{equation}
For $v=3$, the bound is attained by an equilateral triangle.
For $v\ge4$ with $v$ divisible by four, the bound is attained by assigning the inputs equally among the four vertices of a regular tetrahedron, which gives $c_{\min}=1/3$.
\end{lemma}

The proof is given in Appendix~\ref{app:cmin_bound}.
Lemma~\ref{lem:cmin_bound} suggests the regular tetrahedron as the natural geometry for constructing constellations with large $c_{\min}$.
When $v$ is not divisible by four, however, equal multiplicities at the four vertices are impossible, and repeating the vertices of a regular tetrahedron no longer satisfies the balance condition.

We therefore slightly deform the tetrahedron to satisfy the balance condition.
For $v=4m+1$ and $v=4m+3$, one vertex lies on the symmetry axis while the other three are placed symmetrically at a common latitude.
For $v=4m+2$, the four vertices are arranged as two symmetric pairs at different latitudes.
The parameters in the following definition are chosen so that the resulting Bloch vectors satisfy the balance condition.

\begin{definition}[Weighted tetrahedral constellation]\label{def:weighted_constellation}
For $v\ge3$, write
\begin{equation}
    v=4m+t,\qquad t\in\{0,1,2,3\}.
\end{equation}
Define
\begin{equation}
    \kappa(x):=1+((x-1)\bmod 4),\qquad x\in[v],
\end{equation}
and assign
\begin{equation}
    \mathbf r_x:=\mathbf u_{\kappa(x)}.
\end{equation}
The number of input symbols assigned to $\mathbf u_j$ is
\begin{equation}
    \nu_j:=|\{x\in[v]:\kappa(x)=j\}|=m+\mathbf 1\{j\le t\}.
\end{equation}
The four Bloch vectors $\{\mathbf u_j\}_{j=1}^4$ are defined as follows.

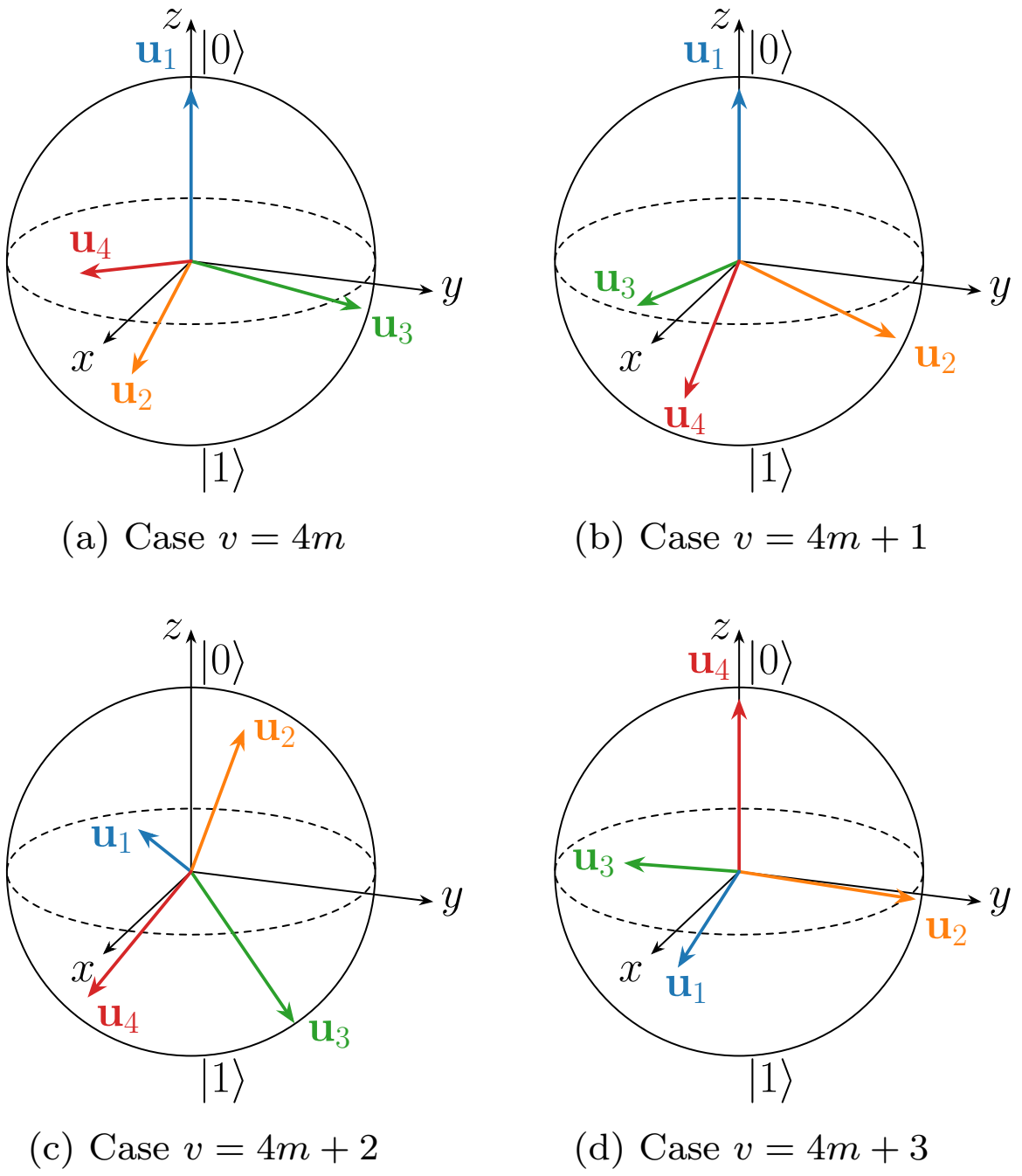
\begin{figure}[t]
    \centering
    \subfloat[Case $v=4m$]{\resizebox{0.48\columnwidth}{!}{
\definecolor{v1}{RGB}{31,119,180}
\definecolor{v2}{RGB}{255,127,14}
\definecolor{v3}{RGB}{44,160,44}
\definecolor{v4}{RGB}{214,39,40}
\newcommand{\drawvec}[6]{
    \draw[->, #1, line width=1.5pt] (0,0,0) -- (#2,#3,#4) node[#5, text=#1] {#6};
}

\tdplotsetmaincoords{70}{110}

\begin{tikzpicture}[scale=3, font=\huge, tdplot_main_coords, >=Stealth, line cap=round, line join=round]
    \draw[tdplot_screen_coords, thick] (0,0) circle (1);
    \tdplotdrawarc[dashed, thick]{(0,0,0)}{1}{0}{360}{}{}
    
    \draw[->, thick] (0,0,0) -- (1.4,0,0) node[below left]{$x$};
    \draw[->, thick] (0,0,0) -- (0,1.4,0) node[right]{$y$};
    \draw[->, thick] (0,0,0) -- (0,0,1.4) node[left]{$z$};
    
    \node[above right, yshift=3pt] at (0,0,1) {$|0\rangle$};
    \node[below right, yshift=-3pt] at (0,0,-1) {$|1\rangle$};
    \drawvec{v1}{0}{0}{1}{above left, xshift = -2pt, yshift = 5pt}{$\mathbf{u}_1$} 
    \drawvec{v2}{0.943}{0}{-0.333}{below}{$\mathbf{u}_2$}
    \drawvec{v3}{-0.471}{0.816}{-0.333}{below right}{$\mathbf{u}_3$}
    \drawvec{v4}{-0.471}{-0.816}{-0.333}{above , xshift=5pt, yshift = 3pt}{$\mathbf{u}_4$} 
\end{tikzpicture}}}
    \subfloat[Case $v=4m+1$]{\resizebox{0.48\columnwidth}{!}{
\definecolor{v1}{RGB}{31,119,180}
\definecolor{v2}{RGB}{255,127,14}
\definecolor{v3}{RGB}{44,160,44}
\definecolor{v4}{RGB}{214,39,40}
\newcommand{\drawvec}[6]{
    \draw[->, #1, line width=1.5pt] (0,0,0) -- (#2,#3,#4) node[#5, text=#1] {#6};
}

\tdplotsetmaincoords{70}{110}

\begin{tikzpicture}[scale=3, font=\huge, tdplot_main_coords, >=Stealth, line cap=round, line join=round]
    \draw[tdplot_screen_coords, thick] (0,0) circle (1);
    \tdplotdrawarc[dashed, thick]{(0,0,0)}{1}{0}{360}{}{}
    
    \draw[->, thick] (0,0,0) -- (1.4,0,0) node[below left]{$x$};
    \draw[->, thick] (0,0,0) -- (0,1.4,0) node[right]{$y$};
    \draw[->, thick] (0,0,0) -- (0,0,1.4) node[left]{$z$};
    
    \node[above right, yshift=3pt] at (0,0,1) {$|0\rangle$};
    \node[below right, yshift=-3pt] at (0,0,-1) {$|1\rangle$};
    \drawvec{v1}{0}{0}{1}{above left, xshift = -2pt, yshift = 5pt}{$\mathbf{u}_1$}
    \drawvec{v2}{-0.433}{0.75}{-0.5}{below right, xshift = 4pt}{$\mathbf{u}_2$}
    \drawvec{v3}{-0.433}{-0.75}{-0.5}{above left, xshift=4pt}{$\mathbf{u}_3$} 
    \drawvec{v4}{0.866}{0}{-0.5}{below}{$\mathbf{u}_4$}
\end{tikzpicture}}}
    \\
    \subfloat[Case $v=4m+2$]{\resizebox{0.48\columnwidth}{!}{
\definecolor{v1}{RGB}{31,119,180}
\definecolor{v2}{RGB}{255,127,14}
\definecolor{v3}{RGB}{44,160,44}
\definecolor{v4}{RGB}{214,39,40}
\newcommand{\drawvec}[6]{
    \draw[->, #1, line width=1.5pt] (0,0,0) -- (#2,#3,#4) node[#5, text=#1] {#6};
}

\tdplotsetmaincoords{70}{110}

\begin{tikzpicture}[scale=3,font=\huge,  tdplot_main_coords, >=Stealth, line cap=round, line join=round]
    \draw[tdplot_screen_coords, thick] (0,0) circle (1);
    \tdplotdrawarc[dashed, thick]{(0,0,0)}{1}{0}{360}{}{}
    
    \draw[->, thick] (0,0,0) -- (1.4,0,0) node[below left]{$x$};
    \draw[->, thick] (0,0,0) -- (0,1.4,0) node[right]{$y$};
    \draw[->, thick] (0,0,0) -- (0,0,1.4) node[left]{$z$};
    
    \node[above right, yshift=3pt] at (0,0,1) {$|0\rangle$};
    \node[below right, yshift=-3pt] at (0,0,-1) {$|1\rangle$};
    \drawvec{v1}{0.845}{0}{0.535}{below left,  xshift = 2pt,yshift=5pt}{$\mathbf{u}_1$}
    \drawvec{v2}{-0.845}{0}{0.535}{right, yshift=-2pt}{$\mathbf{u}_2$}
    \drawvec{v3}{0}{0.598}{-0.802}{right, xshift = 2pt, yshift = -5pt}{$\mathbf{u}_3$}
    \drawvec{v4}{0}{-0.598}{-0.802}{below right}{$\mathbf{u}_4$}
\end{tikzpicture}}}
    \subfloat[Case $v=4m+3$]{\resizebox{0.48\columnwidth}{!}{
\definecolor{v1}{RGB}{31,119,180}
\definecolor{v2}{RGB}{255,127,14}
\definecolor{v3}{RGB}{44,160,44}
\definecolor{v4}{RGB}{214,39,40}
\newcommand{\drawvec}[6]{
    \draw[->, #1, line width=1.5pt] (0,0,0) -- (#2,#3,#4) node[#5, text=#1] {#6};
}

\tdplotsetmaincoords{70}{110}

\begin{tikzpicture}[scale=3,font=\huge,  tdplot_main_coords, >=Stealth, line cap=round, line join=round]
    \draw[tdplot_screen_coords, thick] (0,0) circle (1);
    \tdplotdrawarc[dashed, thick]{(0,0,0)}{1}{0}{360}{}{}
    
    \draw[->, thick] (0,0,0) -- (1.4,0,0) node[below left]{$x$};
    \draw[->, thick] (0,0,0) -- (0,1.4,0) node[right]{$y$};
    \draw[->, thick] (0,0,0) -- (0,0,1.4) node[left]{$z$};
    
    \node[above right, yshift=3pt] at (0,0,1) {$|0\rangle$};
    \node[below right, yshift=-3pt] at (0,0,-1) {$|1\rangle$};
    \drawvec{v4}{0}{0}{1}{above left, yshift=5pt}{$\mathbf{u}_4$}
    \drawvec{v2}{-0.488}{0.844}{-0.222}{below right, yshift=-5pt}{$\mathbf{u}_2$}
    \drawvec{v3}{-0.488}{-0.844}{-0.222}{left}{$\mathbf{u}_3$}
    \drawvec{v1}{0.975}{0}{-0.222}{below right, xshift=-10pt}{$\mathbf{u}_1$}
\end{tikzpicture}}}
    \caption{Weighted tetrahedral Bloch-vector constellations for the four congruence classes of the alphabet size.}
    \label{fig:constellation}
\end{figure}

\paragraph*{Case $v=4m$:}
Let $\{\mathbf u_1,\mathbf u_2,\mathbf u_3,\mathbf u_4\}$ be the vertices of a regular tetrahedron inscribed in the Bloch sphere, so that
\begin{equation}
    \mathbf{u}_{j} \cdot \mathbf{u}_{j'}=-\frac{1}{3},\qquad j \ne j'.
\end{equation}

\paragraph*{Case $v=4m+1$:}
Here $m\ge1$. Let
\begin{equation}
    a:=-\frac{m+1}{3m},\qquad R:=\sqrt{1-a^2}.
\end{equation}
Set
\begin{equation}
    \mathbf u_1:=(0,0,1).
\end{equation}
For $j\in\{2,3,4\}$, let $\theta_j:=2\pi(j-2)/3$ and define
\begin{equation}
    \mathbf u_j:=\left(R\cos\theta_j,R\sin\theta_j,a\right).
\end{equation}

\paragraph*{Case $v=4m+2$:}
Here $m\ge1$. Let
\begin{equation}
    \xi:=\sqrt{\frac{m}{3m+1}},\qquad
    \zeta:=\frac{m+1}{\sqrt{m(3m+1)}}.
\end{equation}
For $j \in \{1,2 \}$, define
\begin{align}
    \mathbf u_j&:=\left((-1)^{j-1}\sqrt{1-\xi^2},0,\xi\right), \nonumber\\
    \mathbf u_{j+2}&:=\left(0,(-1)^{j-1}\sqrt{1-\zeta^2},-\zeta\right).
\end{align}

\paragraph*{Case $v=4m+3$:}
Let
\begin{equation}
    a:=-\frac{m}{3(m+1)},\qquad R:=\sqrt{1-a^2}.
\end{equation}
Set
\begin{equation}
    \mathbf u_4:=(0,0,1).
\end{equation}
For $j\in\{1,2,3\}$, let $\theta_j:=2\pi(j-1)/3$ and define
\begin{equation}
    \mathbf u_j:=\left(R\cos\theta_j,R\sin\theta_j,a\right).
\end{equation}

In every case,
\begin{equation}
    \sum_{x=1}^v\mathbf r_x=\sum_{j=1}^4\nu_j\mathbf u_j=0.
\end{equation}
We denote the resulting balanced constellation by $\mathcal R_v^{\mathrm{prop}}$ and define
\begin{equation}
    \Gamma_v^{\mathrm{prop}}:=\Gamma_{\mathcal R_v^{\mathrm{prop}}},
    \qquad
    \Gamma_v^{\mathrm{perm,prop}}:=\Gamma_{\mathcal R_v^{\mathrm{prop}}}^{\mathrm{perm}}.
\end{equation}
\end{definition}

Figure~\ref{fig:constellation} illustrates the four constellations. We next evaluate the privacy-relevant quantity $c_{\min}$ for these constellations.

\begin{lemma}\label{lem:c_min_calculation}
For the weighted tetrahedral constellation $\mathcal R_v^{\mathrm{prop}}$ in Definition~\ref{def:weighted_constellation},
\begin{equation}
\setlength{\jot}{2pt}
    c_{\min}=\left\{
\begin{alignedat}{2}
        &\dfrac13,&\quad&\text{if }v\equiv0\pmod4,\\
        &\dfrac{2m-1}{6m},&\quad&\text{if }v\equiv1\pmod4,\\
        &\dfrac{m}{3m+1},&\quad&\text{if }v\equiv2\pmod4,\\
        &\dfrac{4m^2+6m+3}{12(m+1)^2},&\quad&\text{if }v\equiv3\pmod4.
\end{alignedat}
\right.
\end{equation}
\end{lemma}

The proof is given in Appendix~\ref{app:cmin_calc}.
 \subsubsection{Comparison with Optimal Classical LDP Mechanisms}\label{sec:comparison}
We now compare the error exponents achieved by the proposed protocols with the classical PUTs and show that a strict quantum advantage remains under the restriction $d=k=2$.
Since $\Gamma^{\mathrm{perm,prop}}_v$ uses input-independent shared randomness, we allow the same resource in the classical comparison.
This does not improve the classical PUTs: if each conditional channel $q_g$ is $\epsilon$-LDP, then
\begin{equation}
    \widetilde q(g,y|x):=p_{G}(g)q_g(y|x)
\end{equation}
is itself an $\epsilon$-LDP channel with a finite output alphabet.
This allows us to evaluate the quantum advantage of the proposed EACC protocols with the permutation ensemble against the classical PUT for symmetric hypothesis testing.

For asymmetric hypothesis testing, combining Theorem~\ref{thm:exact_asym} with the corresponding classical PUT in Theorem~\ref{thm:known_optima} gives
\begin{equation}\label{eq:asym_ratio}
    \lim_{\epsilon \to 0} \frac{A_\eta (\Gamma^{\mathrm{prop}}_v)}{A_\eta[\mathfrak M ^{\mathrm{C}}_\epsilon]} =\frac{v^2} {4\lfloor v/2\rfloor\lceil v/2\rceil(1-c_{\min})}.
\end{equation}
Likewise, Theorem~\ref{thm:exact_ens} and the classical PUT for symmetric hypothesis testing give
\begin{multline}
\lim_{\epsilon\to0}\frac{S_\eta (\Gamma_v^{\mathrm{perm,prop}})}{S_\eta[\mathfrak M_\epsilon^{\mathrm C}]}\\
=\frac{v\eta^2}{16(v-1)(1-c_{\min})}\frac{4v(v-1)}{\lfloor v/2\rfloor\lceil v/2\rceil\eta^2}\\
=\frac{v^2}{4\lfloor v/2\rfloor\lceil v/2\rceil(1-c_{\min})}.
\end{multline}
Substituting the values of $c_{\min}$ from Lemma~\ref{lem:c_min_calculation} shows that both limits equal $f(v)$ defined in Eq.~\eqref{eq:f}.
Since the proposed protocols belong to the respective protocol classes, the corresponding PUTs are no smaller
than their achieved error exponents.
The global minimum of $f(v)$ is $5/4$, attained at $v=5$, and $\lim_{v\to\infty}f(v)=3/2$.
This completes the proof of Theorem~\ref{thm:qubit_advantage}.

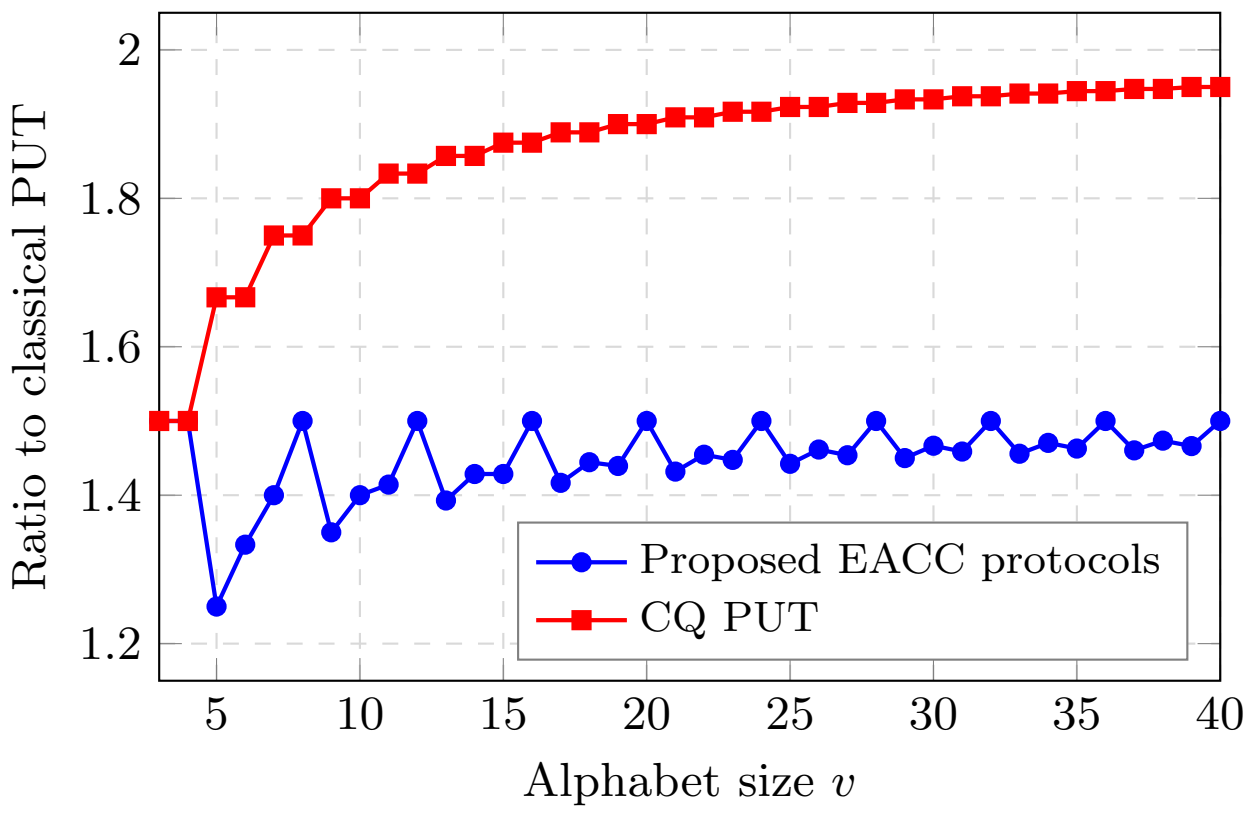
\begin{figure}[tb]
    \centering
    \begin{tikzpicture}[
    declare function={
        qldp(\x) = (mod(\x,2)==0) * (2 - 2/\x) + (mod(\x,2)==1) * (2*\x / (\x+1));
        prop(\x) =
            (mod(\x,4)==0) * (1.5) +
            (mod(\x,4)==1) * (3*\x / (2*\x + 2)) +
            (mod(\x,4)==2) * ((3*\x - 2) / (2*\x)) +
            (mod(\x,4)==3) * (3*\x*(\x+1) / (2*(\x-1)*(\x+3)));
    }
]
    \begin{axis}[
        font=\rmfamily,
        width=8.6cm,  
        height=6cm,
        xlabel={Alphabet size $v$},
        ylabel={Ratio to classical PUT},
        xmin=3, xmax=40,
        ymin=1.15, ymax=2.05,
        xtick={5, 10, 15, 20, 25, 30, 35, 40},
        ytick={1.2, 1.4, 1.6, 1.8, 2.0},
        grid=both,
        grid style={dashed, gray!30},
        legend pos=south east,
        legend cell align={left},
        legend style={
            font=\footnotesize,
            cells={anchor=west},
            draw=black!50
        }
    ]
    \addplot [
        color=blue,
        mark=*,
        mark size=1.5pt,
        thick,
        samples at={3,...,40}
    ] {prop(x)};
    \addlegendentry{Proposed EACC protocols}
    \addplot [
        color=red,
        mark=square*,
        mark size=1.5pt,
        thick,
        samples at={3,...,40}
    ] {qldp(x)};
    \addlegendentry{CQ PUT}
    \end{axis}
\end{tikzpicture}
    \caption{Ratios to the corresponding classical PUTs.
            The proposed EACC protocols achieve the same ratio for asymmetric hypothesis testing and for symmetric hypothesis testing with shared randomness. Both ratios are evaluated as $\epsilon\to0$ and are independent of $\eta\in(0,1]$.}
    \label{fig:advantage_ratio}
\end{figure}

The ratio between the CQ PUTs and classical PUTs in Theorem~\ref{thm:known_optima} converges to $2$ as $v\to\infty$, whereas the ratio between the error exponents achieved by the proposed protocols and the classical PUTs converges to $3/2$ as illustrated in Figure~\ref{fig:advantage_ratio}.
For $v=3$ and $v=4$, the ratio between the error exponents achieved by the proposed protocols and the classical PUTs coincides with the ratio between the CQ PUTs and classical PUTs in Theorem~\ref{thm:known_optima}.
The comparison quantifies how much of the quantum advantage in the EACC-based scenario without resource constraints is retained by the proposed protocols.
It does not assert optimality over $\mathfrak M_{\epsilon,\mathrm{SR}}^{\mathrm{EACC}}(2,2)$ or $\mathfrak M_\epsilon^{\mathrm{EACC}}(2,2)$.

\section{Conclusion} \label{sec:conclusion} 

We investigated whether quantum advantage in
privacy-preserving hypothesis testing can persist
under EACC, where users and the server share entanglement
in advance and exchange only classical messages.
To this end, we introduced MLDP, which protects private
inputs against arbitrary shared quantum states and
server-side measurements. We established that every
QLDP CQ channel admits an MLDP EACC simulation
preserving both asymmetric and symmetric  hypothesis-testing error exponents,
and hence that the two frameworks achieve identical PUTs without resource
constraints.

We then examined how this quantum advantage can
be retained with limited resources. By exploiting
the symmetry of isoclinic mechanisms constructed
from Radon--Hurwitz EITFFs, we showed that the
optimal PUTs can be asymptotically achieved with
only one bit of classical communication, although
the required local entanglement dimension can grow
exponentially with the input alphabet size.
More importantly, we demonstrated that a strict
quantum advantage persists even when each user
shares only a single entangled qubit pair with
the server and transmits one classical bit.
The key idea is to exploit the geometry of qubit
measurements: suitably arranged Bloch vectors
balance the distinguishability needed for hypothesis
testing against the constraints imposed by MLDP.
Using tetrahedral configurations, together with
shared randomness for symmetric hypothesis testing,
we obtained an advantage over the optimal error exponents under the LDP constraint for every $v\ge3$ in the
high-privacy regime. The achievable ratio is at
least $5/4$ and approaches $3/2$ as $v$ increases.

An important direction for future work is to
characterize the optimal EACC PUTs under fixed
local dimension and classical message-alphabet
size, thereby clarifying the tradeoff between
quantum and classical communication resources.
Other directions include quantifying the role
of shared randomness and investigating whether
multipartite entanglement shared across users
and the server can further improve the achievable
error exponents.

\begin{acknowledgments}
The authors used OpenAI's ChatGPT (GPT-6) to assist in identifying relevant literature for Lemma~\ref{lem:unitary_average_simulation}, deriving the bound in Lemma~\ref{lem:cmin_bound}, and suggesting the permutation-ensemble approach.
ChatGPT was also used to generate code for figures based on author-specified mathematical expressions and visualization requirements, check mathematical derivations and proofs, and correct minor linguistic and typographical errors.
Gemini 3.5 Flash-Lite was also used for checking English usage.
The authors independently reviewed the relevant references, verified the mathematical arguments and figures, and take full responsibility for the content of this work.
\end{acknowledgments}
\appendix

\section{Proof of Proposition~\ref{prop:simplified_mldp}}\label{app:mldp_operator}
\begin{proof}
Assume that $\Lambda_{y|x}\le e^\epsilon\Lambda_{y|x'}$. Then $\Delta_{y,x,x'}:=e^\epsilon\Lambda_{y|x'}-\Lambda_{y|x}\ge 0$. For every POVM element $M_z\ge 0$ and state $\rho_{AE}\ge 0$,
\begin{multline}
e^\epsilon P_{\rho, M}(y,z \vert x') -  P_{\rho, M}(y,z \vert x) = \\
    \Tr[(\Delta_{y,x,x'}\otimes M_z)\rho_{AE}]\ge0,
\end{multline}
which is the MLDP inequality.

Conversely, assume MLDP. Choose a one-dimensional auxiliary system $E$, $M_0=I_E$, and $\rho_{AE}=\ketbra{\psi}_A\otimes\ketbra{0}_E$.
Then, $P_{\rho, M}(y,0 \vert x) = \bra{\psi}\Lambda_{y|x}\ket{\psi}$ and
\begin{equation}
    \bra{\psi}\Lambda_{y|x}\ket{\psi}
    \le e^\epsilon\bra{\psi}\Lambda_{y|x'}\ket{\psi}
\end{equation}
for every pure state $\ket{\psi}\in\mathcal H_A$. Hence $\Lambda_{y|x}\le e^\epsilon\Lambda_{y|x'}$.
\end{proof}

\section{Proof of Lemma~\ref{lem:unitary_average_simulation}}\label{app:proof_unitary_average_simulation}
\begin{proof}
Without loss of generality, let
\begin{equation}
    \tau=\sum_{i=0}^{d'-1}\lambda_i\ketbra{i},
\end{equation}
where $d'\le d$ and $\lambda_i>0$.
Define
\begin{equation}
    \Pi_\tau:=\sum_{i=0}^{d'-1}\ketbra{i},
    \qquad
    \widetilde\tau^{-1/2}
    :=\sum_{i=0}^{d'-1}\frac1{\sqrt{\lambda_i}}\ketbra{i},
\end{equation}
where $\Pi_\tau$ is the orthogonal projection onto $\operatorname{supp}\tau$ and
$\widetilde\tau^{-1/2}$ is the inverse square root of $\tau$ on its support.
Then
\begin{equation}
    \widetilde\tau^{-1/2}\tau\widetilde\tau^{-1/2}
    =\Pi_\tau,
    \;
    \tau^{1/2}\widetilde\tau^{-1/2}
    =
    \widetilde\tau^{-1/2}\tau^{1/2}
    =
    \Pi_\tau.
\end{equation}

Since each term in Eq.~\eqref{eq:unitary_average} is positive semidefinite,
\begin{equation}
    0\le U_y^\dagger Q_xU_y\le k\tau,
    \qquad
    \forall x\in\mathcal X,\ y\in\mathcal Y,
\end{equation}
so every $U_y^\dagger Q_xU_y$ is supported on $\operatorname{supp}\tau$.

Let $\mathcal H_A\simeq\mathcal H_S\simeq\mathcal H_B$.
The user and the server share $\sigma_{AS}:=\ketbra{\psi_\tau}_{AS}$, where $\ket{\psi_\tau}_{AS}:= \sqrt d (I_A\otimes\tau^{1/2})\ket{\Phi_d^+}_{AS}$.
This is a valid state because $\Tr\tau=1$, and it is independent of the private input.

Define
\begin{equation}\label{eq:unitary_average_povm}
    \Lambda_{y|x}
    :=
    \frac1k
    \left(
        \widetilde\tau^{-1/2}
        U_y^\dagger Q_xU_y
        \widetilde\tau^{-1/2}
    \right)^T
    +
    \frac1k(I_A-\Pi_\tau^T).
\end{equation}
Equation~\eqref{eq:unitary_average} implies
\begin{equation}
    \sum_{y\in\mathcal Y}\Lambda_{y|x}=I_A,
\end{equation}
so $\{\Lambda_{y|x}\}_{y\in\mathcal Y}$ is a POVM for every $x$.

For any operator $M$ on $A$, the maximally entangled state satisfies
\begin{equation}\label{eq:transpose_mes}
    \Tr_A[(M\otimes I_S)\Phi_{AS}^d]
    =\frac1dM^T.
\end{equation}
Using this identity, the subnormalized server state defined in Eq.~\eqref{eq:rho_yx} is
\begin{align}
    \rho_{y|x}(\Gamma)
    &=
    \Tr_A[
        (\Lambda_{y|x}\otimes I_S)
        \sigma_{AS}
    ]
    \nonumber\\
    &=
    \frac1kU_y^\dagger Q_xU_y.
\end{align}

Define the server-side decoding map
\begin{equation}\label{eq:lemma_decoder}
    \mathcal T(\xi)
    :=
    \sum_{y\in\mathcal Y}
    U_y
    (\bra{y}\otimes I_S)
    \xi
    (\ket{y}\otimes I_S)
    U_y^\dagger.
\end{equation}
Its Kraus operators are
\begin{equation}
    V_y:=U_y(\bra{y}\otimes I_S),
\end{equation}
which satisfy
\begin{equation}
    \sum_yV_y^\dagger V_y
    =
    I_{\mathbb C^k}\otimes I_{S}.
\end{equation}
Hence $\mathcal T$ is CPTP.
Conditioned on the received message, this map applies the corresponding unitary to the server's quantum system and discards the message register.
Therefore,
\begin{equation}
    \mathcal T(\widehat Q_x(\Gamma))
    =
    \sum_{y\in\mathcal Y}
    U_y\rho_{y|x}(\Gamma)U_y^\dagger
    =
    Q_x.
\end{equation}
Thus, $\Gamma$ simulates the CQ channel $Q$ exactly.

Finally, if $Q$ is $\epsilon$-QLDP, then
\begin{multline}
    e^\epsilon\Lambda_{y|x'}
    -
    \Lambda_{y|x}
    =
    \frac1k
    \left[
        \widetilde\tau^{-1/2}
        U_y^\dagger
        (e^\epsilon Q_{x'}-Q_x)
        U_y
        \widetilde\tau^{-1/2}
    \right]^T\\
    +
    \frac{e^\epsilon-1}{k}
    (I_A-\Pi_\tau^T)
    \ge0.
\end{multline}
Proposition~\ref{prop:simplified_mldp} therefore implies that $\Gamma$ is $\epsilon$-MLDP.
\end{proof}
\section{Effective Measurement and Decoding Map of Generalized Quantum Teleportation}\label{app:eff_gen_quant_tel}

The user and the server share a maximally entangled state $\Phi_{AS}^d$, with $\dim\mathcal H_A=\dim\mathcal H_S=d$. 
To derive the effective POVM, consider the standard teleportation procedure in which the user prepares $Q_x$ on an auxiliary $d$-dimensional system $A'$.
The user then performs the generalized Bell measurement $\{\ketbra{\Psi_{ab}}\}_{a,b}$ jointly on systems $A'$ and $A$, where $\ket {\Psi_{ab}} := (I_{A'} \otimes W_{ab}) \ket{\Phi^+_d} $ for $a,b \in\{0,\ldots, d-1\}$ and $W_{ab}$ are the Heisenberg--Weyl operators. We label the outcome $(a,b)$ by $y:=ad+b+1$. The EACC output before correction is
\begin{align}
    \widehat Q_x &= \sum_{a, b} \ketbra{y}{y} \otimes \Tr_{A'A} \left[ (\ket {\Psi_{ab}}\bra{\Psi_{ab}} \otimes I_S)(Q_x \otimes \Phi_{AS}^d) \right] \nonumber\\&
    = \sum_{a, b} \ketbra{y}{y} \otimes  
    \Tr_A\left[\left(\Lambda_{ab|x}^{\mathrm{tel}}\otimes I_S\right)\Phi_{AS}^d\right],
\end{align}
where $\Lambda_{ab|x}^{\mathrm{tel}} :=\Tr_{A'}\left[(Q_x\otimes I_A)\ket{\Psi_{ab}}\bra{\Psi_{ab}}\right]$.

Since $(M\otimes I) \ket {\Phi_d^+} = (I \otimes M^T )\ket{\Phi_d^+}$,
\begin{align}
 \Lambda_{ab|x}^{\mathrm{tel}} &= W_{ab}\Tr_{A'}\left[(Q_x\otimes I_A)\Phi_{A'A}^d\right] W_{ab}^\dagger \nonumber
 \\&= \frac 1d W_{ab} Q_x^T  W_{ab}^\dagger.
\end{align}
Then, 
\begin{align}
 \widehat Q_x & = \sum_{a, b} \ketbra{y}{y} \otimes  \Tr_A\left[\left( \Lambda_{ab|x}^{\mathrm{tel}} \otimes I_S\right)\Phi_{AS}^d\right] \nonumber\\ 
 & = \frac1d\sum_{a, b} \ketbra{y}{y} \otimes (\Lambda_{ab|x}^{\mathrm{tel}})^T\nonumber\\
 &= \frac 1{d^2}\sum_{a, b} \ketbra{y}{y} \otimes  W_{ab}^*Q_xW_{ab}^T.
\end{align}
Note that $\Lambda_{ab|x}^{\mathrm{tel}}$ is positive semidefinite since $Q_x$ is positive semidefinite. Applying the depolarizing identity~\cite{werner2001teleportation} to $Q_x^T$ gives  $\sum_{a,b}\Lambda_{ab|x}^{\mathrm{tel}}=I_A$. 
Therefore, $\{\Lambda_{ab|x}^{\mathrm{tel}}\}_{a,b}$ is the effective POVM the user performs during quantum teleportation.
From Eq.~\eqref{eq:lemma_decoder}, the server-side decoding map is given by
\begin{equation}
    \mathcal T(\xi)=\sum_{a,b}W_{ab}^T(\bra{y}\otimes I_S)\xi(\ket{y}\otimes I_S)W_{ab}^*.
\end{equation}
This map satisfies $\mathcal T(\widehat Q_x)=Q_x$ for every $x\in\mathcal X$.

To show that the protocol also preserves the error exponents,
define the input-independent map
\begin{equation}\label{eq:teleportation_reverse_map}
    \mathcal E(\xi) :=\frac1{d^2}\sum_{a,b} \ketbra{y}\otimes W_{ab}^*\xi W_{ab}^T,
\end{equation}
where $y=ad+b+1$ as above.
Its Kraus operators are
\begin{equation}
    K_{ab}:=\frac1d\ket{y}\otimes W_{ab}^*,
\end{equation}
which satisfy
\begin{equation}
    \sum_{a,b}K_{ab}^\dagger K_{ab}
    =\frac1{d^2}\sum_{a,b}W_{ab}^TW_{ab}^*
    =I_d.
\end{equation}
Hence $\mathcal E$ is CPTP.
Moreover, we can check that $\mathcal E(Q_x)=\widehat Q_x$ for every $x\in\mathcal X$.
Thus, for the constructed protocol $\Gamma$,
\begin{equation}
    \widehat Q(\Gamma)=\mathcal E\circ Q,
    \qquad
    Q=\mathcal T\circ\widehat Q(\Gamma).
\end{equation}

\section{Privacy and Error-Exponent Derivations}\label{app:privacy_derivations}

We first record identities for qubit operators that will be used in the proofs below.

\begin{lemma}\label{lem:qubit_spectral_identities}
Let $a\in\mathbb R$ and $\mathbf b\in\mathbb R^3$. The Hermitian operator
\begin{equation}
    H=aI_2+\mathbf b\cdot\bm\sigma
\end{equation}
has eigenvalues $a\pm\|\mathbf b\|$. In addition, for unit vectors $\mathbf u,\mathbf v\in\mathbb R^3$, define
\begin{equation}
    P_\pm(\mathbf u):=\frac12(I_2\pm\mathbf u\cdot\bm\sigma).
\end{equation}
Then we have
\begin{align}
    \Tr[P_\pm(\mathbf u)P_\pm(\mathbf v)]
    &=\frac{1+\mathbf u\cdot\mathbf v}{2},\\
    \Tr[P_\pm(\mathbf u)P_\mp(\mathbf v)]
    &=\frac{1-\mathbf u\cdot\mathbf v}{2}.
\end{align}
\end{lemma}

\begin{proof}
The identity
\begin{equation}
    (\mathbf u\cdot\bm\sigma)(\mathbf v\cdot\bm\sigma)
    =(\mathbf u\cdot\mathbf v)I_2+i(\mathbf u\times\mathbf v)\cdot\bm\sigma
\end{equation}
gives $(\mathbf b\cdot\bm\sigma)^2=\|\mathbf b\|^2I_2$, which yields the eigenvalues of $H$. The trace identities follow from $\Tr(\sigma_j)=0$ and $\Tr(\sigma_j\sigma_\ell)=2\delta_{j\ell}$.
\end{proof}

\subsection{Proof of Lemma~\ref{lem:exact_privacy}}\label{app:binary_privacy}

By Proposition~\ref{prop:simplified_mldp}, it suffices to impose
\begin{align}
    &e^\epsilon\left(\frac{1-\mu}{2}I_2+\mu P_{x',y}\right)
    -\left(\frac{1-\mu}{2}I_2+\mu P_{x,y}\right)\nonumber\\
    &\quad=\frac{1-\mu}{2}(e^\epsilon-1)I_2
    +\mu\left(e^\epsilon P_{x',y}-P_{x,y}\right)\ge0.
\end{align}
Let
\begin{equation}
    K_{x,x',y}:=e^\epsilon P_{x',y}-P_{x,y}.
\end{equation}
The condition is equivalent to
\begin{equation}
    \frac{1-\mu}{2}(e^\epsilon-1)
    +\mu\lambda_{\min}(K_{x,x',y})\ge0.
\end{equation}
Using the definition of $P_{x,y}$,
\begin{equation}
    K_{x,x',y}
    =\frac{e^\epsilon-1}{2}I_2
    +\frac{(-1)^{y-1}}2(e^\epsilon\mathbf r_{x'}-\mathbf r_x)\cdot\bm\sigma.
\end{equation}
By Lemma~\ref{lem:qubit_spectral_identities},
\begin{equation}
    \lambda_{\min}(K_{x,x',y})
    =\frac{e^\epsilon-1}{2}-\|\mathbf b\|,
\end{equation}
where
\begin{equation}
    \mathbf b:=\frac{(-1)^{y-1}}2 \left(e^\epsilon\mathbf r_{x'}-\mathbf r_x\right).
\end{equation}

Since $\|\mathbf b\|^2 =\frac14\left[(e^\epsilon-1)^2+4e^\epsilon(1-c_{x,x'})\right]$,
\begin{multline}
    \lambda_{\min}(K_{x,x',y})
    =\frac12\left[ e^\epsilon-1 \right.\\\left.
    -\sqrt{(e^\epsilon-1)^2
    +4e^\epsilon(1-c_{x,x'})} \right].
\end{multline}
This expression is increasing in $c_{x,x'}$, so the most restrictive pair has overlap $c_{\min}$. Substituting the above equation into the scalar eigenvalue condition and solving for $\mu$ gives Eq.~\eqref{eq:mu_privacy_bound}.

For the choice of $\mu$ that meets Eq.~\eqref{eq:mu_privacy_bound} with equality,
\begin{equation}\label{eq:mu_high_privacy_limit}
    \frac{\mu^2}{\epsilon^2}
    =\frac{[(e^\epsilon-1)/\epsilon]^2}
    {(e^\epsilon-1)^2+4e^\epsilon(1-c_{\min})}
    \longrightarrow\frac1{4(1-c_{\min})},
\end{equation}
as $\epsilon\to0$.

\subsection{Proof of Theorem~\ref{thm:exact_asym}}\label{app:asym_proof}

By Eq.~\eqref{eq:asym_stein},
\begin{equation}
    A_\eta\left(\Gamma_{\mathcal R}\right)
    =\min_{h\in[v]}D\left(\omega_{\eta, h}(\Gamma_{\mathcal R})\Vert\omega_0(\Gamma_{\mathcal R})\right).
\end{equation}

By the direct-sum property of relative entropy~\cite[Exercise 11.8.8]{wilde2017}, the block-diagonal structure in Eq.~\eqref{eq:binary_server_state} gives
\begin{multline}
    D\left(\omega_{\eta, h}(\Gamma_{\mathcal R})\Vert\omega_0(\Gamma_{\mathcal R})\right)
    \\=\frac{1}{2}\sum_{y=1}^2
    D\left(\varphi_{y|\eta, h}(\Gamma_\mathcal R)\middle\|\frac14I_2\right).
\end{multline}

By Lemma~\ref{lem:qubit_spectral_identities}, the eigenvalues of $\varphi_{y|\eta, h}$ are
\begin{equation}
    \lambda_\pm=\frac{1\pm\mu\eta}{4},
\end{equation}
independently of $h$ and $y$. Hence
\begin{align}
    D\left(\varphi_{y|\eta, h} (\Gamma_\mathcal R)\middle\|\frac14I_2\right)
    &=\frac{1+\mu\eta}{2}\log(1+\mu\eta)\nonumber\\
    &\quad+\frac{1-\mu\eta}{2}\log(1-\mu\eta).
\end{align}
The right-hand side is independent of $y$, so averaging over the two message values gives the expression stated in the theorem.

Let
\begin{equation}
    h(t):=\frac{1+t}{4}\log(1+t)+\frac{1-t}{4}\log(1-t).
\end{equation}
Since $h(t)=t^2/4+\mathcal O(t^4)$ as $t \to 0$,
\begin{equation}
    A_\eta \left(\Gamma_{\mathcal R}\right) = 2h(\mu\eta)
    =\frac{\mu^2\eta^2}{2}+\mathcal O(\mu^4).
\end{equation}
Combining this expansion with Eq.~\eqref{eq:mu_high_privacy_limit} proves the high-privacy limit.

\subsection{Proof of Theorem~\ref{thm:exact_sym}}\label{app:sym_proof}

For $h\ne h'$, define
\begin{equation}
    C_{h,h'}(s):=\Tr\left[\omega_{\eta, h}(\Gamma_{\mathcal R})^s\omega_{\eta, h'}(\Gamma_{\mathcal R})^{1-s}\right].
\end{equation}
From the block-diagonal structure of $\omega_{\eta, h}$,
\begin{multline}
    C_{h,h'}(s)\\=\sum_{y=1}^2\Tr\left((\varphi_{y|\eta, h}(\Gamma_\mathcal R))^s(\varphi_{y|\eta, h'}(\Gamma_\mathcal R))^{1-s}\right).
\end{multline}

By Lemma~\ref{lem:qubit_spectral_identities}, the subnormalized states have common eigenvalues $\lambda_\pm=(1\pm\mu\eta)/4$. For $y\in[2]$, using Eq.~\eqref{eq:varphi_yh},
\begin{equation}
    \varphi_{y|\eta, h}(\Gamma_\mathcal R)=\lambda_+P_{h,y}^T+\lambda_-P_{h,3-y}^T.
\end{equation}
Since $P_{h,y}^T$ and $P_{h,3-y}^T$ are mutually orthogonal projections, for $s\in(0,1)$,
\begin{equation}
    \left(\varphi_{y|\eta, h}(\Gamma_\mathcal R)\right)^{s}=\lambda_+^sP_{h,y}^T+\lambda_-^sP_{h,3-y}^T.
\end{equation}
Equation~\eqref{eq:pairwise_overlap} gives
\begin{align}
    \Tr(P_{h,y}^TP_{h',y}^T)=\Tr(P_{h,3-y}^TP_{h',3-y}^T)&=c_{h,h'},\\
    \Tr(P_{h,y}^TP_{h',3-y}^T)=\Tr(P_{h,3-y}^TP_{h',y}^T)&=1-c_{h,h'}.
\end{align}
Therefore,
\begin{multline}
        \Tr\left((\varphi_{y|\eta, h}(\Gamma_\mathcal R))^s(\varphi_{y|\eta, h'}(\Gamma_\mathcal R))^{1-s}\right)\\
    =c_{h,h'}\left(\lambda_+^s\lambda_+^{1-s}+\lambda_-^s\lambda_-^{1-s}\right)\\
    \quad+(1-c_{h,h'})\left(\lambda_+^s\lambda_-^{1-s}+\lambda_-^s\lambda_+^{1-s}\right).
\end{multline}

Since $\lambda_++\lambda_-=1/2$, we obtain
\begin{multline}
    \Tr\left((\varphi_{y|\eta, h}(\Gamma_\mathcal R))^s(\varphi_{y|\eta, h'}(\Gamma_\mathcal R))^{1-s}\right)\\=\frac12c_{h,h'}+(1-c_{h,h'})g(s),
\end{multline}
where
\begin{equation}
    g(s):=\lambda_+^s\lambda_-^{1-s}+\lambda_-^s\lambda_+^{1-s}.
\end{equation}

The expression is the same for both $y\in[2]$, and hence
\begin{equation}
    C_{h,h'}(s)=c_{h,h'}+2(1-c_{h,h'})g(s).
\end{equation}
The function $g$ is convex and symmetric under $s\mapsto1-s$, so $s=1/2$ is a global minimizer. Moreover,
\begin{equation}
    g(1/2)=2\sqrt{\lambda_+\lambda_-}=\frac12\sqrt{1-\mu^2\eta^2}.
\end{equation}
Therefore,
\begin{equation}
    \min_{s\in[0,1]}C_{h,h'}(s)=c_{h,h'}+(1-c_{h,h'})\sqrt{1-\mu^2\eta^2}.
\end{equation}
This quantity is increasing in $c_{h,h'}$, so the smallest pairwise Chernoff exponent is attained at $c_{\max}$. This proves the stated expression.

Finally,
\begin{equation}
    S_\eta \left(\Gamma_{\mathcal R}\right)=\frac{1-c_{\max}}2\mu^2\eta^2+\mathcal O(\mu^4),
\end{equation}
and the limit follows from Eq.~\eqref{eq:mu_high_privacy_limit}.
\section{Permutation-Ensemble Derivations}\label{app:permutation}

\subsection{Proof of Theorem~\ref{thm:exact_ens}}\label{app:ensemble_proof}

For one user, Eq.~\eqref{eq:shared_randomness_state} gives
\begin{equation}
    \!\!\Omega_{\eta, h}
    :=\omega_{\eta, h}^{\mathrm{SR}}(\Gamma_{\mathcal R}^{\mathrm{perm}})
    =\frac1{v!}\sum_{\pi\in S_v}
    \ketbra{\pi}\otimes\omega_{\eta, \pi(h)}(\Gamma_{\mathcal R}).
\end{equation}
Because the shared random permutations are independent across users, the $n$-user observation under hypothesis $h$ is $\Omega_{\eta,h}^{\otimes n}$. For $h\ne h'$,
\begin{multline}
        \Tr(\Omega_{\eta, h}^s\Omega_{\eta, h'}^{1-s})\\=\frac1{v!}\sum_{\pi\in S_v}\Tr\left[\omega_{\eta,\pi(h)}(\Gamma_{\mathcal R})^s\omega_{\eta,\pi(h')}(\Gamma_{\mathcal R})^{1-s}\right].
\end{multline}
Each summand has the convexity and $s\mapsto1-s$ symmetry established in the proof of Theorem~\ref{thm:exact_sym}. Hence the average is minimized at $s=1/2$. At this point,
\begin{align}
    \Tr(\Omega_{\eta,h}^{1/2}\Omega_{\eta, h'}^{1/2})
    &=\frac1{v!}\sum_{\pi\in S_v}
    \left[c_{\pi(h),\pi(h')}\right.\nonumber\\
    &\left.\qquad+(1-c_{\pi(h),\pi(h')})
    \sqrt{1-\mu^2\eta^2}\right].
\end{align}
A uniform permutation maps the ordered pair $(h,h')$ uniformly onto all ordered pairs $(x,x')$ with $x\ne x'$. Therefore,
\begin{equation}
    \Tr(\Omega_{\eta,h}^{1/2}\Omega_{\eta, h'}^{1/2})
    =\bar c+(1-\bar c)\sqrt{1-\mu^2\eta^2},
\end{equation}
independently of the choice of distinct $h$ and $h'$, which proves the stated expression. The high-privacy expansion is
\begin{equation}
    S_{\eta}\left(\Gamma_{\mathcal R}^{\mathrm{perm}}\right)
    =\frac{1-\bar c}{2}\mu^2\eta^2+\mathcal O(\mu^4).
\end{equation}
Using $1-\bar c=v/[2(v-1)]$ and Eq.~\eqref{eq:mu_high_privacy_limit} proves the stated limit.

\section{Geometric Derivations}\label{app:geometry}

\subsection{Proof of Lemma~\ref{lem:cmin_bound}}
\label{app:cmin_bound}
For $v=3$, the balance condition and the unit norms imply $\mathbf r_x\cdot\mathbf r_{x'}=-1/2$ for all $x\ne x'$,
giving $c_{\min}=1/4$.

For $v\ge4$, suppose, for contradiction, that $c_{\min}>1/3$.
Then all pairwise inner products are greater than $-1/3$.
By~\cite[Lem.~4.1]{boroczky2020isodiametric}, there exists a vector $\mathbf u\in\mathbb R^3$ such that $\mathbf u\cdot\mathbf r_x>0$ for every $x\in[v]$.
This contradicts the balance condition
$\sum_x\mathbf r_x=0$.
Hence $c_{\min}\le1/3$.

\subsection{Proof of Lemma~\ref{lem:c_min_calculation}}\label{app:cmin_calc}

Since the overlap is affine in the Bloch inner product, it suffices to determine
\begin{equation}
    s_{\min}:=\min_{x\ne x'}\mathbf r_x\cdot\mathbf r_{x'}
\end{equation}
for each congruence class $v=4m+t$.

\paragraph*{Case $v=4m$:}
For all distinct $j,j'\in\{1,2,3,4\}$,
\begin{equation}
    \mathbf u_j\cdot\mathbf u_{j'}=-\frac13.
\end{equation}
Therefore,
\begin{equation}
    s_{\min}=-\frac13,
    \qquad
    c_{\min}=\frac13.
\end{equation}

\paragraph*{Case $v=4m+1$:}
For $j\in\{2,3,4\}$,
\begin{equation}
    \mathbf u_1\cdot\mathbf u_j=a=-\frac{m+1}{3m},
\end{equation}
whereas, for distinct $j,j'\in\{2,3,4\}$,
\begin{equation}
    \mathbf u_j\cdot\mathbf u_{j'}=\frac{3a^2-1}{2}.
\end{equation}
Since $a<(3a^2-1)/2$ for $m\ge1$,
\begin{equation}
    s_{\min}=-\frac{m+1}{3m},
    \qquad
    c_{\min}=\frac{2m-1}{6m}.
\end{equation}

\paragraph*{Case $v=4m+2$:}
For $j\in\{1,2\}$ and $j'\in\{3,4\}$,
\begin{equation}
    \mathbf u_j\cdot\mathbf u_{j'}
    =-\xi\zeta=-\frac{m+1}{3m+1}.
\end{equation}
In addition,
\begin{align}
    \mathbf u_1\cdot\mathbf u_2
    &=2\xi^2-1=-\frac{m+1}{3m+1},\\
    \mathbf u_3\cdot\mathbf u_4
    &=2\zeta^2-1\ge-\frac{m+1}{3m+1}.
\end{align}
Hence
\begin{equation}
    s_{\min}=-\frac{m+1}{3m+1},
    \qquad
    c_{\min}=\frac{m}{3m+1}.
\end{equation}

\paragraph*{Case $v=4m+3$:}
For distinct $j,j'\in\{1,2,3\}$,
\begin{equation}
    \mathbf u_j\cdot\mathbf u_{j'}
    =\frac{3a^2-1}{2}
    =\frac{-2m^2-6m-3}{6(m+1)^2}.
\end{equation}
When $m\ge1$, symbols are also assigned to $\mathbf u_4$, and for $j\in\{1,2, 3\}$,

\begin{equation}
    \mathbf u_4\cdot\mathbf u_j=a=-\frac{m}{3(m+1)}.
\end{equation}

The first value is no greater than $a$. When $m=0$, the set assigned to $\mathbf u_4$ is empty, and only the pairs among $\mathbf u_1,\mathbf u_2,\mathbf u_3$ occur. 
Therefore, in all cases,
\begin{equation}
    s_{\min}=\frac{-2m^2-6m-3}{6(m+1)^2}, \;
    c_{\min}=\frac{4m^2+6m+3}{12(m+1)^2}.
\end{equation}
This completes the proof.

\bibliography{ref}
\end{document}